\pdfoutput=1

\documentclass[10pt]{article}

\usepackage{listings}    % for snippets of code using \begin{lstlisting}.
\usepackage{graphicx}    % for figures, eg, \includegraphics
\usepackage{subcaption}  % for subfigures, ie figures with internal figures
\usepackage{algorithm}   % for displaying algorithm in pseudo-code, using \begin{algorithm}
\usepackage{algorithmic} % and \begin{algorithmic}
\usepackage{booktabs}    % for better tables, eg using \toprule
\usepackage{hyperref}    % to handle URL links with \url
\usepackage[htt]{hyphenat} % for line breaks in texttt, using \-
\usepackage{microtype} % ligatures are extremely annoying, especially when copy&paste text from PDF
\usepackage[algo2e,ruled,vlined,linesnumbered]{algorithm2e}
\DisableLigatures{}    % https://tex.stackexchange.com/questions/439651/how-do-i-disable-ligatures

\usepackage{caption}
\usepackage{setspace}
\usepackage{multirow}
\usepackage{enumerate}
\usepackage{pdflscape}
\usepackage{pgfplots} % bar lib

\usepackage{xcolor}
\definecolor{codegreen}{rgb}{0.25,0.5,0.35}
\definecolor{codegray}{rgb}{0.5,0.5,0.5}
\definecolor{codepurple}{rgb}{0.6,0,0}
\definecolor{backcolour}{rgb}{0.95,0.95,0.92}
\definecolor{colorstring}{rgb}{0.5,0,0.35}
\definecolor{rltred}{rgb}{0.5,0,0}
\definecolor{rltgreen}{rgb}{0,0.5,0}
\definecolor{rltblue}{rgb}{0,0,0.5}
\definecolor{DarkGreen}{rgb}{0.00,0.60,0.00}
\definecolor{ScarletRed}{rgb}{0.80,0.00,0.00}
\definecolor{blizzardblue}{rgb}{0.67, 0.9, 0.93}
\definecolor{green-yellow}{rgb}{0.68, 1.0, 0.18}
\definecolor{dkgreen}{rgb}{0,0.6,0}
\definecolor{gray}{rgb}{0.5,0.5,0.5}
\definecolor{mauve}{rgb}{0.58,0,0.82}
\definecolor{lightgrey}{rgb}{0.90,0.90,0.90}
\definecolor{grey}{gray}{0.75}
\definecolor{light-gray}{gray}{0.80}

\lstdefinestyle{mystyle}{
    escapechar=©, %  use ©\label{}© when needing \label pointing to line numbers
	backgroundcolor=\color{backcolour},
    basicstyle=\footnotesize\ttfamily,
   	identifierstyle=\footnotesize\ttfamily,
	commentstyle=\color{codegreen},
	keywordstyle=\color{colorstring}\bfseries,
	numberstyle=\ttfamily\color{codegray},
	stringstyle=\ttfamily\color{DarkGreen},
	breakatwhitespace=false,
	breaklines=true,
	captionpos=b,
	keepspaces=true,
	numbers=left, % possible values are (none, left, right)
	numbersep=2pt,
	showspaces=false,
	showstringspaces=false,
	showtabs=false,
	tabsize=2
}
\usepackage{xspace}
\newcommand{\evo}{{\sc EvoMaster}\xspace}
\newcommand{\etal}{{\emph{et al.}}\xspace}

\usepackage{boxedminipage}
\newenvironment{result}%
{\smallskip
	\noindent
	\let\emph=\textbf
	\begin{boxedminipage}{\columnwidth}\begin{center}\em}%
		{\end{center}\end{boxedminipage}%
}

\usepackage{ifthen}
\newboolean{showcomments}
\setboolean{showcomments}{true} % comment this line to deactivate comments

\ifthenelse{\boolean{showcomments}}{
	\newcommand{\nbc}[3]{
		{\colorbox{#3}{\bfseries\sffamily\scriptsize\textcolor{white}{#1}}}
		{\textcolor{#3}{\sf\small$\langle$\textit{#2}$\rangle$}}}
	
}{
	\newcommand{\nbc}[3]{}

}

\newcommand{\rpcncs}{{\emph{rpc-thrift-ncs}}\xspace}

\newif\ifSpringer
\Springerfalse

\newif\ifArxiv
\Arxivfalse

\usepackage{authblk}                % for \affil
\usepackage[square,numbers]{natbib} % needed for the references, when using ACM-Reference-Format outside an
\usepackage{amsmath}                % for math operations

\usepackage{amsfonts}

\usepackage{geometry}
\title{
Causes and Effects of Fitness Landscapes in System Test Generation: A Replication Study
}

\author[1]{Omur Sahin}
\author[2]{Man Zhang}
\author[3,4]{Andrea Arcuri}

\affil[1]{Erciyes University, Kayseri, Türkiye}
\affil[2]{Beihang University, Beijing, China}
\affil[3]{Kristiania University College, Oslo, Norway}
\affil[4]{Oslo Metropolitan University, Oslo, Norway}

\date{}
\Arxivtrue
\begin{document}

\maketitle

\begin{abstract}
Search-Based Software Testing (SBST) has seen several success stories in academia and industry.
The effectiveness of a search algorithm at solving a software engineering problem strongly depends on how such algorithm can navigate the \emph{fitness landscape} of the addressed problem.
The fitness landscape depends on the used fitness function.
Understanding the properties of a fitness landscape can help to provide insight on how a search algorithm behaves on it.
Such insight can provide valuable information to researchers to being able to design novel, more effective search algorithms and fitness functions tailored for a specific problem.
Due to its importance, few fitness landscape analyses have been carried out in the scientific literature of SBST.
However, those have been focusing on the problem of \emph{unit test} generation, e.g., with state-of-the-art tools such as EvoSuite.
In this paper, we \emph{replicate} one such existing study.
However, in our work we focus on \emph{system test} generation, with the state-of-the-art tool \evo.
Based on an empirical study involving the testing of 23 web services, this enables us to provide valuable insight into this important testing domain of practical industrial relevance.
\end{abstract}


\newcommand{\totalSuts}{{$23$}\xspace}
\newcommand{\totalRunSuts}{{$23$}\xspace}
\newcommand{\totalBranches}{{$9037$}\xspace}
\newcommand{\totalNeverCovered}{{$2008$}\xspace}
\newcommand{\totalEndpoints}{{$616$}\xspace}
\newcommand{\totalLOCs}{{$471010$}\xspace}
\newcommand{\totalSourceFiles}{{$4704$}\xspace}
\newcommand{\numberOfGraphQL}{{$5$}\xspace}
\newcommand{\numberOfThrift}{{$2$}\xspace}
\newcommand{\numberOfREST}{{$16$}\xspace}
\newcommand{\totalBranchesInFigures}{{$7029$}\xspace}
\newcommand{\baseAlgorithm}{RW}
\newcommand{\compareAlgorithm}{MIO}
\newcommand{\correlationAC}{$-0.58$}\xspace
\newcommand{\correlationDBI}{$0.60$}\xspace
\newcommand{\correlationNV}{$0.61$}\xspace
\newcommand{\correlationND}{$-0.58$}\xspace
\newcommand{\correlationIC}{$0.60$}\xspace
\newcommand{\correlationPIC}{$0.59$}\xspace

\newcommand{\baseAlgorithmText}{\baseAlgorithm\xspace}
\newcommand{\compareAlgorithmText}{\compareAlgorithm\xspace}

% file names
\newcommand{\landscapeFigFile}{generated_files/figs/landscape}
\newcommand{\sutsTableFile}{generated_files/tables/suts-info}
\newcommand{\branchInfoTableFile}{generated_files/tables/branches-info}
\newcommand{\comparisonTableFile}{generated_files/tables/comparisons_\compareAlgorithm_\baseAlgorithm}
\newcommand{\sixMeasureTableFile}{generated_files/tables/six-measure_\baseAlgorithm}
\newcommand{\sixMeasureGroupTableFile}{generated_files/tables/six-measure-group_\baseAlgorithm}
\newcommand{\neverCoveredTable}{generated_files/tables/never-covered-branch-types_\baseAlgorithm_\compareAlgorithm}
\newcommand{\figureOneFile}{generated_files/figs/Figure1_\baseAlgorithm_\compareAlgorithm}
\newcommand{\figureOneHistogramFile}{generated_files/figs/Figure1_\baseAlgorithm_\compareAlgorithm_histogram}
\newcommand{\figureTwoFile}{generated_files/figs/Figure2_\baseAlgorithm_\compareAlgorithm}
\newcommand{\figureThreeFile}{generated_files/figs/Figure3_\baseAlgorithm_\compareAlgorithm}
\newcommand{\figureSevenFile}{generated_files/figs/Figure7_\baseAlgorithm_\compareAlgorithm}
\newcommand{\figureEightFile}{generated_files/figs/Figure8_\baseAlgorithm_\compareAlgorithm}

\newcommand{\branchTypesTable}{generated_files/tables/branch-types}

\definecolor{OliveGreen}{cmyk}{0.64,0,0.95,0.40}
\newcommand\omur[1]{\nbc{Omur}{#1}{OliveGreen}}

\newcommand{\rqA}{{What are the characteristics of the fitness landscape for system-level test case generation?}\xspace}
\newcommand{\rqB}{{How do the fitness landscape characteristics, like neutrality and ruggedness, influence search algorithms in test generation?}\xspace}
\newcommand{\rqC}{{How do source code attributes, like complex preconditions and boolean variables, affect the fitness landscape and the effectiveness of search-based testing?}\xspace}
\newcommand{\rqD}{{What are the differences in how the characteristics of the fitness landscape affect search-based testing in unit test generation versus system-level test generation?}\xspace}

%%%%%%%%%%%%%%%%%%%%%%%%%%%%%%%%%%%%%%%%%%%%%%%%%%%%%%%%%%%%%%%%%%%%%%%%%%%%
\section{Introduction}
Many modern applications are built using web services such as REST~(\cite{fielding2000architectural}), SOAP~(\cite{curbera2002unraveling}), or GraphQL~(\cite{quina2023graphql}).
In large and complex enterprise applications, these are structured into individual web services through a microservice architecture~(\cite{newman2021building}).
This method minimizes the development and maintenance costs associated with monolithic applications and aims to create more robust solutions.
Major companies such as Netflix, Uber, eBay, Amazon, and Nike have widely adopted this approach in the industry~(\cite{rajesh2016spring}).

However, testing web services presents many challenges due to their complexity~(\cite{bozkurt2013testing, canfora2009service}).
In particular for REST APIs, several techniques have been proposed in the literature~(\cite{golmohammadi2023testing}).
Various tools have been proposed for fuzzing web services in recent years, including
ARAT-RL~(\cite{kim2023adaptive}),
bBOXRT~(\cite{laranjeiro2021black}),
DeepREST~(\cite{corradini2024deeprest}),
\evo~(\cite{arcuri2025tool}),
Morest~(\cite{liu2022icse}),
ResTest~(\cite{martinLopez2021Restest}),
RestCT~(\cite{wu2022icse}),
Restler~(\cite{restlerICSE2019}),
RestTestGen~(\cite{viglianisi2020resttestgen})
and
Schemathesis~(\cite{hatfield2022deriving}).
Apart from \evo, all these tools are \emph{black-box}, unable to analyze the source code of the tested APIs to achieve better results.
Although black-box testing has its place in industry~(\cite{icst2025vw}), results with \emph{white-box} testing, when applicable, can be much better~(\cite{zhang2023open}).

A key component for white-box \evo, that enables it to achieve significantly better results, is the use of
Search-Based Software Testing (SBST) techniques,
whose applicability and effectiveness have been demonstrated in various studies~(\cite{ABHP09, evosuiteAtSbst2013, harman2012search, mao2016sapienz}).
In SBST, software test generation is cast into a search problem,
%The search action is performed
which can then been tackled with various search algorithms, such as
Genetic Algorithm (GA)~(\cite{holland1992genetic}),
Many Objective Sorting Algorithm (MOSA)~(\cite{dynamosa2017}),
or Many Independent Objective (MIO)~(\cite{mio2017}).
 These algorithms are guided by a \emph{fitness function} to try to find the best solution in the \emph{search space} of all possible test cases for the given problem.

Several studies in the literature of SBST demonstrate that these algorithms can effectively help test case generation while the fitness functions accurately direct the algorithm.
However, existing fitness functions are not able to guide the algorithms in covering all test targets, such as covering all branches~(\cite{campos2018empirical, albunian2020causes}).
In addition to familiar challenges, such as complex parameters that complicate covering all targets during a search, the lack of a comprehensive understanding of search behavior further complicates the identification of the factors that contribute to these difficulties~(\cite{albunian2020causes}).
One method used to understand search behavior is \emph{fitness landscape analysis}~(\cite{zou2022survey}), which helps to understand the difficulties underlying the problems.
A deep understanding of the problem's search space enhances insight into algorithm behavior and helps in improving algorithms' problem-solving capabilities.

The two primary properties of fitness landscapes that significantly impact the optimization process are \emph{ruggedness} and \emph{neutrality}~(\cite{malan2013survey}).
The interaction between these properties has inspired the creation of various techniques to analyze their structure.
The main purpose of this study is to explore the characteristics of the search space involved in system-level test generation by analyzing the fitness landscape.
To achieve this, we measured six different proxy metrics commonly used to assess system tests' ruggedness and neutrality properties. These calculations were carried out using the \evo tool to perform a random walk in the search space.
Results and analyses were conducted on a total of \totalRunSuts Systems Under Test (SUTs) featuring REST, GraphQL, and RPC applications.

This work aims at addressing the following research questions:
\begin{description}
    \item[{\bf RQ1}:] \rqA
    \item[{\bf RQ2}:] \rqB
    \item[{\bf RQ3}:] \rqC
    \item[{\bf RQ4}:] \rqD
\end{description}

This work is a \emph{replication study} of a fitness landscape analysis of \emph{unit test generation}~(\cite{albunian2020causes}).
However, in this study the difference is that we aim at analyzing the fitness landscape of \emph{system-level test generation} to evaluate the complexity of branches.
This research aims at improving SBST approaches, by collecting insights into the impact of neutrality factors, such as challenging preconditions, and boolean flags.
The key research questions address how these neutrality factors influence the search process and how enhancing fitness functions or \emph{testability transformations}~(\cite{HHH04,arcuri2021tt}) can mitigate these challenges.
This study employs the MIO algorithm to determine branch difficulty, followed by a Random Walk analysis for deeper insights.
This work contributes to automated test generation methods to improve software reliability and maintainability.

The paper is organized as follows.
Section~\ref{sec:background} provides background information on SBST, on \evo and on the used algorithms in our study.
Related work is discussed in Section~\ref{sec:relatedwork}.
Section~\ref{sec:fitnesslandscape} presents the details of our replicated fitness landscape analysis.
Empirical analyses are presented in Section~\ref{sec:study}.
Our observations on these results follow in Section~\ref{sec:discussion}.
Threats to validity are discussed in Section~\ref{sec:threats}.
Finally, Section~\ref{sec:conclusions} concludes the paper.

%RQ1: What are the properties of the fitness landnscape for the system-level test case generation problem?
%RQ2: How do the fitness landscape properties affect the search behavior?
%RQ3: What are the underlying properties of source code that influence the fitness landscape?

%%%%%%%%%%%%%%%%%%%%%%%%%%%%%%%%%%%%%%%%%%%%%%%%%%%%%%%%%%%%%%%%%%%%%%%%%%%%%%%%%%%%%%%%%%%%%%%%%%%%%%%%%%%%%%%%%%%%
\section{Background}
\label{sec:background}

%--------------------------------------------------------------------------------------------------------------------
\subsection{Search-Based Software Testing}
Testing is a crucial but challenging aspect of software development, often seen by developers as a tedious task, particularly for complex applications.
Furthermore, besides being possibly tedious and expensive, \emph{manual} testing can lack rigor and not be systematic, which can lead to poor verification effectiveness.
For these reasons, lot of research has been carried out on how to test software \emph{automatically}~(\cite{bertolino2007software}).

A common approach to ease this process is to use \emph{random} test case generation~(\cite{DuN84,AIB11}), which is suitable for simple cases but typically fails to provide effective coverage for complex scenarios.
Search-Based Software Testing (SBST) approaches have been developed to address these complex scenarios effectively.
In SBST, software testing is cast into an optimization problem, which can then be tackled with search algorithms (e.g., Genetic Algorithms~(\cite{holland1992genetic})).
Many applications using the SBST approach have been developed (e.g., (\cite{Mcm04,HaJ01,harman2012search,ABHP09})).
Tools like EvoSuite~(\cite{fraser2011evosuite}) and Pynguin~(\cite{lukasczyk2022pynguin}) for unit testing, Sapienz~(\cite{mao2016sapienz}) for mobile testing, and EvoMaster~(\cite{arcuri2025tool}) for Web API testing use search algorithms to systematically explore input domains and maximize objectives like code coverage and fault detection.

The optimization process in SBST is driven by a \emph{fitness function} that evaluates test cases based on how well they achieve testing goals.
Techniques like evolutionary algorithms are commonly applied, where operators such as \emph{crossover} and \emph{mutator} help refine test cases over successive generations, improving their ability to explore the software’s behavior.
This evolutionary approach allows SBST to navigate the \emph{fitness landscape}, effectively identifying test cases for complex and hidden faults in the code.

To achieve better results, various heuristics are used to ``smooth'' the search landscape, providing guidance to the search algorithm in finding an optimal solution.
In the context of white-box testing, a well-known technique introduced in the 1990s by Korel~(\cite{Kor90}) is called \emph{Branch Distance}.
This technique was initially developed to address predicates involving numerical comparisons (e.g., $a < b$) and was later refined to handle logical operators~(\cite{gallagher1997adtest}) (e.g., AND and OR) as well as string comparisons~(\cite{AlB06}).

The branch distance enables the application of optimization methods by providing a smoother fitness landscape.
The code snippet shown in Figure~\ref{fig:example_code} is divided into two branches.
If we consider here the ``then'' branch related to satisfying the constraint \texttt{x==42}, the branch distance can be defined as $d(x)=|x-42|$.
So, the code snippet has a landscape as given in Figure~\ref{fig:landscape}, and the algorithm performs its search in this search space of integer inputs.
This is a rather straightforward search landscape.
Small modifications to the inputs (e.g., via mutation operator) have clear gradient (i.e., fitness scores improve) towards the \emph{global optimum}, without any risk of being stuck in \emph{local optima} or in \emph{fitness plateaus}.

However, in the general case, generating test cases can be complex due to combinatorial challenges, making the fitness landscape more intricate.
To address this complexity, we use proxy methods for fitness landscape analysis, as explained in Section~\ref{sec:fitnesslandscape}.

%--------------------
\begin{figure}[!t]
    \begin{lstlisting}[language=java,basicstyle=\footnotesize]
public static void foo(int x){
    if(x == 42){
        print("The answer to the ultimate question of life, the universe, and everything");
    } else {
        print("Wrong. Try again");
    }
}
    \end{lstlisting}
    \caption{Code snippet of a simple numerical function with a single \texttt{if} statement.} \label{fig:example_code}
\end{figure}
%--------------------
\begin{figure}[t]
    \centering
    \includegraphics[width=.6\linewidth]{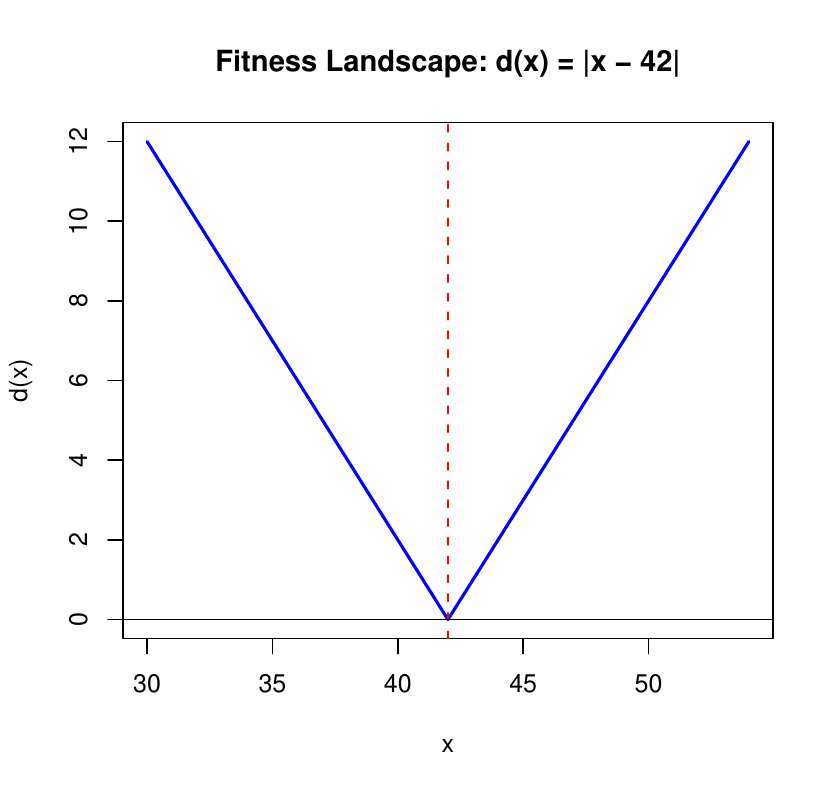}
    %\vspace{-2.5\baselineskip}
    \caption{Fitness landscape of the code snippet in Figure~\ref{fig:example_code}.}
    \label{fig:landscape}
    %\vspace{-1.0\baselineskip}
\end{figure}
%--------------------

%---------------------------------------------------------------------------------------------
\subsection{EvoMaster}
\label{sec:evomaster}

\evo~(\cite{arcuri2025tool,arcuri2021evomaster,arcuri2019restful}) is an open-source fuzzer for automated Web API fuzzing that supports black~(\cite{arcuri2020blackbox}) and white-box~(\cite{arcuri2020testability}) testing modes.
Black-box testing can be applied on APIs written in any programming language.
For white-box testing, \evo is currently compatible with APIs developed in Java and Kotlin~(\cite{arcuri2019restful}),
although in the past there was support for JavaScript/TypeScript~(\cite{zhang2023javascript}) and C\#~(\cite{golmohammadi2023net}) (which are no longer maintained).

\evo generates test cases for Web APIs using evolutionary algorithms.
It contains search methods, including the Many Independent Objective (MIO)~(\cite{arcuri2018test}) algorithm (which is set as the default), the Whole Test Suite (WTS)~(\cite{GoA_TSE12}), and the Many-Objective Sorting Algorithm (MOSA)~(\cite{dynamosa2017}).
 Additionally, to enhance API fuzzing, \evo has been improved with techniques such as testability transformations~(\cite{arcuri2021enhancing}), SQL handling~(\cite{arcuri2020sql}), adaptive hypermutation~(\cite{zhang2021adaptive}) and few others advanced white-box heuristics~(\cite{arcuri2023advanced}).

\evo provides automated fuzzing for various types of APIs, including REST~(\cite{arcuri2019restful,zhang2021resource,zhang2021enhancing}), GraphQL~(\cite{belhadi2023random}), and RPC~(\cite{zhang2023rpc}) APIs.
Notably, it is the only fuzzer for Web APIs in the literature that supports white-box testing, using an automated method to capture runtime data such as SBST heuristics via bytecode manipulation (using the same approach as older SBST tools such as EvoSuite).
%As RPC API fuzzing, \evo offers extensive support for frameworks like gRPC~\cite{gRPC} and Apache Thrift~\cite{zhang2023rpc}.
\evo is a mature tool, downloaded thousands of times~(\cite{arcuri2025tool}) and used for example in Fortune 500 enterprises such as Meituan~(\cite{zhang2023rpc,zhang2024seeding}) and Volkswagen~(\cite{poth2025technology,icst2025vw}).

Due to all these features, in this study we selected \evo as the main tool for our fitness landscape analyses in the context of system test generation.

%---------------------------------------------------------------------
\subsection{The RW Algorithm}
\label{sub:rw}
One essential tool for fitness landscape analysis is the concept of a walk~(\cite{pitzer2012comprehensive}). Suppose we visualize a landscape similar to a real-world setting. In that case, the basic principle is to continually move a short distance from one solution candidate to a nearby one while closely monitoring the fitness progress. There are various methods such as random walk (RW), adaptive walk, reverse adaptive walk, uphill-down
and neutral walk~(\cite{pitzer2012comprehensive}).
RW was used in this study. The RW algorithm is widely used for analyzing landscapes in large and complex problems. It starts from a randomly chosen point and makes random steps from there. The pseudo-code of the RW algorithm used in this study is given in Algorithm~\ref{algo:rw}.

First, an individual is sampled randomly. Then, the same individual is mutated until the stopping criterion is met, and if there is an improvement in any target, it is added to the archive. Here, the stopping criterion is number of steps, i.e., the number of individual evaluations.
\begin{algorithm2e}
	\DontPrintSemicolon
	\SetAlgoLined
	\SetKwInOut{Input}{Input}\SetKwInOut{Output}{Output}
	\Input{Stopping condition $C$, Fitness function $\delta$, Population size $n$}
	\Output{Archive of optimized individuals $A$}
	\BlankLine
	$p \leftarrow null$ \;
	\While{$\neg$$C$}{
		\eIf{$p == null$}{
			$p \leftarrow RandomIndividual(\,)\,$\;
		}{
			$p \leftarrow Mutate(p)\,$\;
		}
		\ForEach{element $k$ $\in$ ReachedTargets($p$)}{
			\If{NewTarget(k)}{
				$S \leftarrow S \cup T_{k}$\;
			}
			$T_{k} \leftarrow T_{k}\cup \{p\}$\;
			\uIf{$IsTargetCovered(k)\,$}{
				$UpdateArchive(A, p)\,$\;
				$S\leftarrow S \setminus T_{k}$\;
			}
			\ElseIf{$|T_{k}| > n$}{
				$RemoveWorst(T_{k}, \delta)\,$\;
			}
		}
	}
	
	\caption{Pseudo-code of the RW Algorithm}\label{algo:rw}
\end{algorithm2e}

%--------------------------------------------------------
\subsection{The MIO Algorithm}
\label{sub:mio}
The MIO algorithm~(\cite{mio2017,arcuri2018test}) is designed specifically for white-box system-level test generation.
It is the default search algorithm in \evo.
The pseudo-code can be found in Algorithm~\ref{algo:mio}. 
MIO is a genetic-based evolutionary algorithm inspired by the (1+1) Evolutionary Algorithm (EA)~(\cite{DJW02}).
This section will provide a brief discussion, but full details can be found in~(\cite{mio2017,arcuri2018test}).

MIO is a multi-population algorithm designed for multiple test targets, with a separate population for each target. 
At the beginning of the search process, a random population is initialized based on a chromosome template using information obtained from the schema of the tested API.

During each iteration, MIO samples a target either randomly or by selecting from the population of targets that have not yet been covered. 
A test case consists of one or more actions representing HTTP calls used to test a web service. 
A new solution is generated by applying only the mutation operator. 
Two types of mutations can be applied: structural mutation and internal mutation. 
Structural mutation changes the structure of the test case by adding or removing actions, while internal mutation changes the values of the genes of an action, such as flipping a boolean value between \texttt{true} and \texttt{false}.

The fitness value of each new test is calculated after it is sampled or mutated. 
If an improvement is observed in any target, it is recorded in the corresponding population, and the worst individuals are removed. 
When a test successfully covers a target, this individual is added to an archive. 
The relevant population is then reduced to one individual and no longer used for new sampling.
At the end of the search, a test suite consisting of the best tests is generated.

\begin{algorithm2e}
	\DontPrintSemicolon
	\SetAlgoLined
	\SetKwInOut{Input}{Input}\SetKwInOut{Output}{Output}
	\Input{Stopping condition $C$, Fitness function $\delta$, Population size $n$, Probability for random sampling $P_{r}$, Start of focused search $F$}
	\Output{Archive of optimized individuals $A$}
	\BlankLine
	$S \leftarrow SetOfEmptyPopulations(\, )\,$ \;
	$A \leftarrow \{\}$ \;
	\While{$\neg$$C$}{
		\eIf{$P_{r} > rand(\,)\,$}{
			$p \leftarrow RandomIndividual(\,)\,$\;
		}{
			$p \leftarrow SampleIndividual(S)\,$\;
			$p \leftarrow Mutate(p)\,$\;
		}
		\ForEach{element $k$ $\in$ ReachedTargets($p$)}{
			\If{NewTarget(k)}{
				$S \leftarrow S \cup T_{k}$\;
			}
			$T_{k} \leftarrow T_{k}\cup \{p\}$\;
			\uIf{$IsTargetCovered(k)\,$}{
				$UpdateArchive(A, p)\,$\;
				$S\leftarrow S \setminus T_{k}$\;
			}
			\ElseIf{$|T_{k}| > n$}{
				$RemoveWorst(T_{k}, \delta)\,$\;
			}
		}
		$UpdateParameters(F, P_{r}, n)\,$
	}
	
	\caption{Pseudo-code of the MIO Algorithm~\cite{arcuri2018test}}\label{algo:mio}
\end{algorithm2e}

%%%%%%%%%%%%%%%%%%%%%%%%%%%%%%%%%%%%%%%%%%%%%%%%%%%%%%%%%%%%%%%%%%%%%%%%%%%%
\section{Related Work}
\label{sec:relatedwork}

Regarding \emph{fitness landscape analyses} in SBST, there are a few existing works.
Waeselynck \etal~(\cite{WFK06}) analyze the search space's ruggedness and size using a diameter metric to configure a simulated annealing algorithm for test generation.
Lefticaru and Ipate~(\cite{lefticaru2008comparative}) explore the local structure and size of the search space, similar to Waeselynck \etal~(\cite{WFK06}).
They introduce a problem hardness measure called fitness distance correlation, which requires knowledge of the global optimum. Their analysis also evaluates various fitness functions for specification-based testing.
Aleti \etal~(\cite{aleti2017analysing}) analyze the fitness landscape for generating test suites with EvoSuite using three metrics: population information content, negative slope coefficient, and change rate, primarily to assess solvability.
Albunian \etal~(\cite{albunian2020causes}) examine the landscape's ruggedness and neutrality to generate tests with the Many-Objective Sorting Algorithm (MOSA) using EvoSuite.
Vogel \etal conducted two different studies~(\cite{vogel2019does,vogel2021comprehensive}) and improved the performance of their algorithms by conducting a fitness landscape analysis on Sapienz, which is used for mobile application testing.
Besides software testing, in software engineering Aleti and Moser~(\cite{aleti2015fitness}) address the challenge of optimizing software architectures by focusing on the analysis of local structure, specifically ruggedness.

Another area of research examines the fitness landscape to determine whether it is elementary, or to construct elementary landscapes~(\cite{LBY10,chicano2011elementary}).
Elementary landscapes represent a specific category of fitness landscapes that can guide the development of effective heuristics. These heuristics may be applicable to problems such as the next release problem~\cite{LBY10} or test suite minimization~(\cite{chicano2011elementary}).

Regarding \emph{replication studies} in SBST, those are rare.
Perhaps the most known example is the work on parameter tuning in SBST~(\cite{arcuri2013parameter}), that was replicated in at least two studies~(\cite{sayyad2013parameter,kotelyanskii2014parameter}).
Other examples involve software effort estimation~(\cite{tawosi2021multi})
and tool evolution~(\cite{golmohammadi2023impact}).

%%%%%%%%%%%%%%%%%%%%%%%%%%%%%%%%%%%%%%%%%%%%%%%%%%%%%%%%%%%%%%%%%%%%%%%%%%%%
\section{Fitness Landscape Analysis}
\label{sec:fitnesslandscape}
The fitness function represents the objectives of an optimization problem in evolutionary computing and swarm intelligence algorithms. Fitness values are compared based on the principles of maximization or minimization during iterative processes, and these values are calculated according to the fitness function of the specific problem. Therefore, if the characteristic structure of an optimization problem can be understood in depth, the relationship of this problem to the algorithm behavior can also be understood more easily. Thus, this deep understanding can also help the algorithms' ability to solve challenges effectively~(\cite{zou2022survey}).

If we define the fitness landscape formally~(\cite{albunian2020causes,zou2022survey}), it can be represented as $(X, N, f)$, where $X$ denotes the set of potentially feasible solutions, and $N$ is the neighborhood operator on $X$ (e.g., a mutation operator in evolutionary algorithms).
$f$ is a fitness function ($f: X \rightarrow \mathbb{R}$) that maps each genotype to a numerical fitness value.
A fitness landscape is characterized by various features~(\cite{malan2013survey}).
Among these features, \emph{ruggedness} and \emph{neutrality} significantly influence the ability to find optimal solutions.

\emph{Ruggedness} is one of the key characteristics of a fitness landscape.
If a fitness landscape contains multiple local optima along with a single, isolated global optimum, and the fitness values of neighboring individuals are less correlated, it is considered ``rugged''.
In this scenario, finding the optimal solution can be challenging because an algorithm may become trapped in local optima.

Another important characteristic to consider is \emph{neutrality}.
When there are generally equal values throughout the search space, meaning the search surface consists of plateaus, ruggedness alone may not be sufficiently descriptive.
In these situations, the fitness value can remain constant for extended periods of the search.
When a mutation occurs in a neutral fitness landscape, it can lead to a change in position on the fitness map without affecting the fitness value.
In this case, a neighboring solution of $y$ is called a neutral neighbor if $f(x) = f(y)$ at point $y$.

In this study, we calculate six metrics, including \emph{Autocorrelation (AC)}, \emph{Neutrality Distance (ND)}, \emph{Neutrality Volume (NV)}, \emph{Information Content (IC)}, \emph{Partial Information Content (PIC)}, and \emph{Density-basin Information (DBI)}.
These are the same metrics used in the study we replicate~(\cite{albunian2020causes}), where they were employed to assess the ruggedness and neutrality characteristics of the fitness landscape in unit test generation.
These six metrics will be presented and briefly discussed in the following subsections.
For more in-depth details on these metrics (including their motivation and justification), the interested reader is referred to~ \cite{vassilev2000information} and \cite{pitzer2012comprehensive}.

To calculate these metrics, as done in the literature, there is the need to use a \emph{random walk}.
In a random walk, a solution from $X$ is randomly chosen (i.e., random starting point), and then the neighboring operator $N$ (e.g., the mutation operator in \evo) is applied up to $k$ times (e.g., $k=1000$), each time computing the fitness scores with $f$, and keeping track of all of them (as needed to compute those six metrics).

An additional calculation is required to calculate the DBI, IC and PIC metrics~(\cite{vassilev2000information}) to obtain more advanced information. 
This measurement is calculated using fitness value sequences. 
However, instead of directly using a fitness value, a string expression is generated through a transformation, and the calculations are then performed based on this expression. 
Initially, each sequence of fitness values ($\left\{f_{t}\right\}_{t=1}^{k}$) are transformed into a series of fitness changes using Eq.~\ref{eq:neutrality_volume_one}.
\begin{equation}
	\label{eq:neutrality_volume_one}
	\Delta\left\{f_{t}\right\}_{t=1}^{k}:=\left\{f_{t}-f_{t-1}\right\}_{t=2}^{k}
\end{equation}
Next, a string expression is generated according to Eq.~\ref{eq:neutrality_volume_two}. In this context, the string expression can be defined as $S(\epsilon) = s_1, s_2, s_3, \dots, s_k$, where each expression is denoted as $s_i \in \left\{\overline{1}, 0, 1\right\}$.
\begin{equation}
	\label{eq:neutrality_volume_two}
	s_{i}=\left\{\begin{array}{l l}{{\bar{1},}}
		& {{\mathrm{if~}x\ <-\epsilon}}\\ {{0,}}& {{\mathrm{if~}\left|x\right|\leq\epsilon}}\\ {{1,}}& {{\mathrm{if~}x\ >\epsilon}}
	\end{array}\right.
\end{equation}
where $x$ represents the changes in fitness. The parameter $\epsilon$ is a real number from the interval $[0, l_k]$, and $l_k$ represents the length of the interval of fitness values obtained by the random walk. The $\epsilon$ parameter adjusts the landscape's sensitivity. By changing $\epsilon$, one can “zoom in and out” to observe the same walk with varying levels of detail.

%----------------------------------------------------------------------------
\subsection{Autocorrelation}
The AC is used to measure the correlation between two individuals.
This correlation measurement is calculated for two individuals that differ by a step count of $i$.
The calculation is made using Eq.~\ref{eq:auto_correlation}.
In this equation,
%$N$ 
%represents the total number of individuals in the random walk,
$s$ is the step size, $f_i$ denotes the fitness value of the $i$-th individual, and $\overline{f}$ is the average fitness value of all individuals.
The range of values for $r(s)$ is between $-1$ and $1$. 
%A value closer to $0$ indicates less correlation, meaning the random walk is more rugged. 
% unit_test_landscape.pdf page 4
``\textit{The landscape is more rugged when the AC value is close to $0$ meaning that the individuals of the random walk are less correlated}''~(\cite{albunian2020causes}).

\begin{equation}
    \label{eq:auto_correlation}
    r(s)=\frac{\sum_{i=1}^{k-s}(f_{i}-\overline{f})(f_{i+s}-\overline{f})}{\sum_{i=1}^{k}(f_{i}-\overline{f})^{2}}
\end{equation}

%----------------------------------------------------------------------------
\subsection{Neutrality Distance}
%TODO maybe add a figure to explain the concept
The ND measures neutrality in landscapes by identifying the longest neutral step number during a random walk where no fitness values change. In other words, it represents the largest $t$ value in the equation $f(x_1)=f(x_2)=\dots=f(x_t)$. The range of values for ND is between $0$ and $1$, and it can be calculated using Eq.~\ref{eq:neutral_distance}. The landscape is more neutral when the ND value is close to 1.

\begin{equation}
	\label{eq:neutral_distance}
	ND=\frac{t}{k}
\end{equation}

%----------------------------------------------------------------------------
\subsection{Neutrality Volume}
\label{sub:NV}

The NV is one of the metrics used to measure neutrality. It is calculated based on the number of areas with equal fitness values during a random walk. For instance, if we have a fitness value sequence $f_t = {0.3, 0.3, 0.3, 0.2, 0.2, 0.7, 0.7}$, we can identify $z=3$ distinct regions corresponding to the fitness values of $0.3$, $0.2$, and $0.7$. The NV value is calculated by dividing the number of distinct regions $z$ by the total number of steps $k$, i.e., $NV=z/k$. The value ranges from $0$ to $1$, and neutrality increases as it approaches $0$.

\subsection{Information Content}
\label{sub:IC}

The IC metric is determined by analyzing the diversity within the string $S(\epsilon)$ to assess the ruggedness of the landscape. It is calculated using the entropy of consecutive symbols that differ from each other, based on Eq.~\ref{eq:information_content_one}.
\begin{equation}
    \label{eq:information_content_one}
    {H}(\epsilon)=-\sum_{p\neq q}{P}_{[p q]}\;\log_{6}{P}_{[p q]}
\end{equation}

The probabilities ${P}_{[p q]}$ represent the frequencies of possible blocks $pq$ of elements from the set $[\overline{1}, 0, 1]$, as defined by Eq.~\ref{eq:information_content_two}.

\begin{equation}
    \label{eq:information_content_two}
    {P}_{[p q]}=\frac{n_{[p q]}}{n}
\end{equation}
Where $n_{[pq]}$ refers to the number of times each $pq$ appears in the string $S(\epsilon)$.
The value range of the ${H}(\epsilon)$ is $[0,1]$.
As the number of peaks in the landscape increases (which can imply a higher ruggedness), the value of $H(\epsilon)$ also increases.

%------------------------------------------------------------------------
\subsection{Partial Information Content}
\label{sub:PIC}
The PIC metric is designed to analyze the \emph{modality} in the landscape, which is related to the number of peaks (i.e., local optima) in it.
Modality is often correlated with ruggedness.
To calculate this metric, first, all zero values in $S(\epsilon)$ and all values equal to its preceding symbol are removed, and a new $S'(\epsilon)$ is calculated.
Then, the PIC is calculated using Eq.~\ref{eq:partial_information_content}.
\begin{equation}
    \label{eq:partial_information_content}
    M(\epsilon)=\frac{\mu}{n}
\end{equation}

Where $\mu$ is the length of the $S'$ string, and $n$ is the length of the $S$ string.
The range of values for $M(\epsilon)$ is between $0$ and $1$.
%If the landscape is maximally multimodal, the $M(\epsilon)$ value will be 1, while, if the landscape is flat, it will be 0.
``\textit{If the landscape path is maximally multimodal, $M(\epsilon)$ is $1$ as the string $S'(\epsilon)$ is identical to $S(\epsilon)$ (i.e., $S(\epsilon)$ cannot be modified). 
In contrast, the landscape path is flat when the $M(\epsilon)$ is $0$ as there are no slopes in the landscape path}''~(\cite{albunian2020causes}).

%------------------------------------------------------------------------------
\subsection{Density-basin Information}
\label{sub:DBI}
The DBI method evaluates the variety of flat areas within a landscape by focusing on the characteristics of smooth points. It achieves this by examining consecutive equal symbols in a given string $S$.
In this analysis, the only relevant sub-blocks identified in the string are composed of pairs $[00, 11, \overline{1}\overline{1}]$. This metric can be calculated with Eq.~\ref{eq:density_basin}.
\begin{equation}
    \label{eq:density_basin}
    h(\epsilon)=-\sum_{p=q}P_{[p q]}\log_{3}P_{[p q]}
\end{equation}

The range of values for $h(\epsilon)$ is between $0$ and $1$ and ``\textit{it decreases when the number of groups increases, i.e., the density of peaks becomes lower}''~(\cite{vassilev2000information}).
In other words, high values for DBI typically represent landscapes that are rugged and with low neutrality.

%A high $h(\epsilon)$ value indicates low peak density and suggests that the landscape primarily consists of flat areas.

%-------------------------------------------------------------------------------
\subsection{Summary}

\begin{table}[!t]
	\ifArxiv
	\centering
	\fi
\caption{\label{tab:metrics}
Interpretation for each of the six employed metrics, based on their values.
}
\begin{tabular}{l| c | ccc}
\toprule
Name & Range & Low  & Medium & High \\
\midrule
AC   & $[-1,+1]$  &          & Rugged & \\
ND   & $[0,1]$    &          &        & Neutral \\
NV   & $[0,1]$    & Neutral  &        & \\
IC   & $[0,1]$    &          &        & Rugged \\
PIC  & $[0,1]$    &          &        & Rugged \\
DBI  & $[0,1]$    &          &        & Rugged \\
\bottomrule
\end{tabular}
\end{table}

Each of these six metrics does measure some specific properties of the search landscape.
These properties are all somehow related to the concepts of neutrality and ruggedness.
Considering they have different range values, and different interpretations for the minimal and maximal values,
to clarify them we summarize them in Table~\ref{tab:metrics}.

Recall that \emph{neutrality} is not necessarily the opposite of \emph{ruggedness}.
A landscape could be for example neutral and rugged at the same time.

%%%%%%%%%%%%%%%%%%%%%%%%%%%%%%%%%%%%%%%%%%%%%%%%%%%%%%%%%%%%%%%%%%%%%%%%%%%%
\section{Empirical Study}
\label{sec:study}

To evaluate the fitness landscape characteristics of system-level test case generation, we conducted an empirical study to answer the following research questions:

\begin{description}
	\item[{\bf RQ1}:] \rqA
	\item[{\bf RQ2}:] \rqB
	\item[{\bf RQ3}:] \rqC
	\item[{\bf RQ4}:] \rqD
\end{description}

%----------------------------------------------------------------------
\subsection{Case Study}
\label{sub:case-study}

\begin{table}[!t]
	\ifArxiv
		\centering
	\fi
    \footnotesize
    \caption{Systems Under Test (SUTs) used in our empirical study.} \label{table:sut_info}
    \begin{tabular}{ l l r r r l} 
\toprule 
\emph{SUT} & \emph{Type} & \emph{\#LOCs} & \emph{\#SourceFiles} & \emph{\#Endpoints} & \emph{Language} \\ 
\midrule 
\emph{catwatch} & REST & 9636 & 106 & 14 & Java \\ 
\emph{cwa-verification} & REST & 3955 & 47 & 5 & Java \\ 
\emph{genome-nexus} & REST & 30004 & 405 & 23 & Java \\ 
\emph{gestaohospital-rest} & REST & 3506 & 33 & 20 & Java \\ 
\emph{graphql-ncs} & GraphQL & 548 & 8 & 6 & Kotlin \\ 
\emph{graphql-scs} & GraphQL & 577 & 13 & 11 & Kotlin \\ 
\emph{languagetool} & REST & 174781 & 1385 & 2 & Java \\ 
\emph{market} & REST & 9861 & 124 & 13 & Java \\ 
\emph{ocvn-rest} & REST & 45521 & 526 & 258 & Java \\ 
\emph{patio-api} & GraphQL & 18048 & 178 & 20 & Java \\ 
\emph{pay-publicapi} & REST & 34576 & 377 & 10 & Java \\ 
\emph{petclinic-graphql} & GraphQL & 5212 & 89 & 15 & Java \\ 
\emph{proxyprint} & REST & 8338 & 73 & 74 & Java \\ 
\emph{reservations-api} & REST & 1853 & 39 & 7 & Java \\ 
\emph{rest-ncs} & REST & 605 & 9 & 6 & Java \\ 
\emph{rest-news} & REST & 857 & 11 & 7 & Kotlin \\ 
\emph{rest-scs} & REST & 862 & 13 & 11 & Java \\ 
\emph{restcountries} & REST & 1977 & 24 & 22 & Java \\ 
\emph{rpc-thrift-ncs} & Thrift & 585 & 9 & 6 & Java \\ 
\emph{rpc-thrift-scs} & Thrift & 772 & 14 & 11 & Java \\ 
\emph{scout-api} & REST & 9736 & 93 & 49 & Java \\ 
\emph{session-service} & REST & 1471 & 15 & 8 & Java \\ 
\emph{timbuctoo} & GraphQL & 107729 & 1113 & 18 & Java \\ 
\midrule 
Total & & 471010 & 4704 & 616 &  \\ 
\bottomrule 
\end{tabular} 

\end{table}

In the experiments, we used a total of \totalSuts APIs consisting of REST, GraphQL, and RPC applications present in the EMB corpus~(\cite{icst2023emb}).
EMB is a collection of Web APIs consisting of REST, GraphQL, and RPC APIs, which we have gathered and expanded annually with new additions since 2017.
It also features the~\evo drivers, needed to enable white-box fuzzing for all these APIs.
These drivers are configurations used to specify how to start, stop and reset these SUTs.
They are also responsible to automatically instrument the bytecode of these SUTs when started, to be able to calculate different kinds of SBST heuristics such as the branch distances.

Table~\ref{table:sut_info} shows some statistics on these \totalSuts APIs, including the number of source files, lines of code, and number of endpoints.
This collection comprises \totalRunSuts SUTs: \numberOfREST using REST, \numberOfGraphQL using GraphQL, and~\numberOfThrift using Thrift RPC.
They contain \totalLOCs lines of code across \totalSourceFiles source files, with \totalEndpoints endpoints in total. 
Note that these code statistics reflect only what is contained in the business logic of the APIs.
Data from third-party libraries, like HTTP servers and libraries to access SQL databases, is not accounted for here.

EMB provides APIs of varying sizes and complexities from diverse domains, addressing a broad range of APIs essential for scientific experimentation.
A full description of these APIs can be found at~\cite{icst2023emb}.
This resource includes the source code, build scripts, and links to the original repositories from which these APIs have been gathered over the years.

%----------------------------------------------------------------
\subsection{Experiment Settings}
\label{sub:settings}

To address our research questions, we execute the RW and MIO algorithms.
The RW algorithm (Section~\ref{sub:rw}) employs a mutation operator to explore the landscape randomly,
while the MIO algorithm (Section~\ref{sub:mio}) is used to assess the difficulty of branches.
We follow the same kind of procedure done in~\cite{albunian2020causes}, where MOSA (the default algorithm in EvoSuite) was used to assess the difficulty of the branches.

The RW and MIO algorithms were executed $30$ times each, as done in~\cite{albunian2020causes}, which is a common practice in software engineering research.
The stopping criterion for each run was set at $1000$ steps, indicating $1000$ individual evaluations. 
This is the same number of steps used in~\cite{albunian2020causes}, which is based on common practice in fitness landscape analysis research~(\cite{barnett1998ruggedness}).

Note that, technically, this would not be a fair comparison between algorithms, as each test case can have a different number of HTTP calls (e.g., randomly between 1 and 10 in the default settings of \evo).
However, we are not comparing algorithms (e.g., to show that MIO is better than RW), but rather use them to study the characteristics of the fitness landscape.

The experiments carried out $2$ Configurations $\times$~\totalRunSuts SUTs $\times$ $30$ Runs $\times$ $1000$ Evaluations $= 1380000$ steps, i.e., 1.3M fitness evaluations, with each single step recorded individually for every branch.
These  experiments lasted $\approx106$ hours, generating around 65GB of data to analyze.

The work was carried out with~\evo version 3.4.0, and other control parameters are the default.
The $\epsilon$ parameter for IC (Section~\ref{sub:IC}), PIC (Section~\ref{sub:PIC}), and DBI (Section~\ref{sub:DBI}) used in the analyses was selected as $0$.
Thus, analyses were performed at maximum sensitivity level.

%-----------------------------------------------------------------------------
\subsection{Comparison of MIO and RW}
\label{sub:comparison}

% log_em_rest-scs_17860 -> found potantial 202 faults in rest-scs.
% log_em_catwatch_18330 -> found potantial 2700 faults in catwatch.
\begin{table}[t]
	\ifArxiv
		\centering
	\fi
    \tiny
    \caption{Comparison of the~\compareAlgorithmText and~\baseAlgorithmText algorithms.} \label{table:mio_rw_comparison}
    % Andrea: resize gives issues
    %\resizebox{\textwidth}{!}{
    %\footnotesize
    \begin{tabular}{ l rrrr rrrr rrr}\\ 
\toprule 
SUT & \multicolumn{4}{c}{Line Coverage \%} & \multicolumn{4}{c}{\# Detected Faults} & \multicolumn{3}{c}{\# HTTP Calls} \\ 
    & MIO & RW  & $\hat{A}_{12}$ & p-value  & MIO & RW & $\hat{A}_{12}$ & p-value & MIO & RW & Ratio \\% \\ 
\midrule 
\emph{catwatch} & 43.5 & 34.6 & {\bf 0.12} & $< 0.001$ & 94.4 & 152.1 & {\bf 0.10} & $< 0.001$ & 2541 & 8451 & 332.61\\% \\ 
\emph{cwa-verification} & 43.5 & 36.4 & {\bf 0.28} & 0.003 & 2.4 & 0.8 & {\bf 0.10} & $< 0.001$ & 4343 & 4174 & 96.11\\% \\ 
\emph{genome-nexus} & 34.6 & 28.5 & {\bf 0.15} & $< 0.001$ & 0.0 & 0.0 & 0.50 & 1.000 & 2795 & 2150 & 76.93\\% \\ 
\emph{gestaohospital-rest} & 35.2 & 17.9 & {\bf 0.00} & $< 0.001$ & 20.5 & 9.4 & {\bf 0.05} & $< 0.001$ & 4544 & 2400 & 52.83\\% \\ 
\emph{graphql-ncs} & 82.1 & 50.7 & {\bf 0.00} & $< 0.001$ & 6.2 & 4.6 & {\bf 0.06} & $< 0.001$ & 4975 & 8000 & 160.81\\% \\ 
\emph{graphql-scs} & 69.5 & 49.3 & {\bf 0.00} & $< 0.001$ & 11.0 & 6.0 & {\bf 0.00} & $< 0.001$ & 3123 & 8000 & 256.15\\% \\ 
\emph{languagetool} & 9.3 & 5.4 & {\bf 0.00} & $< 0.001$ & 5.5 & 4.3 & {\bf 0.34} & 0.023 & 1246 & 1340 & 107.47\\% \\ 
\emph{market} & 36.6 & 34.4 & {\bf 0.33} & 0.023 & 44.3 & 24.8 & {\bf 0.25} & $< 0.001$ & 2894 & 7920 & 273.62\\% \\ 
\emph{ocvn-rest} & 20.5 & 14.1 & {\bf 0.00} & $< 0.001$ & 431.4 & 68.0 & {\bf 0.00} & $< 0.001$ & 1484 & 3736 & 251.76\\% \\ 
\emph{patio-api} & 16.2 & 12.3 & {\bf 0.22} & $< 0.001$ & 33.2 & 16.7 & {\bf 0.00} & $< 0.001$ & 4967 & 8000 & 161.05\\% \\ 
\emph{pay-publicapi} & 13.0 & 13.0 & {\bf 0.08} & $< 0.001$ & 41.5 & 24.8 & {\bf 0.00} & $< 0.001$ & 1597 & 2501 & 156.62\\% \\ 
\emph{petclinic-graphql} & 44.2 & 25.7 & {\bf 0.00} & $< 0.001$ & 19.9 & 7.3 & {\bf 0.00} & $< 0.001$ & 5316 & 8000 & 150.49\\% \\ 
\emph{proxyprint} & 27.6 & 9.6 & {\bf 0.00} & $< 0.001$ & 122.0 & 19.2 & {\bf 0.00} & $< 0.001$ & 1693 & 8407 & 496.60\\% \\ 
\emph{reservations-api} & 53.1 & 45.0 & {\bf 0.08} & $< 0.001$ & 106.0 & 7.8 & {\bf 0.00} & $< 0.001$ & 1430 & 1773 & 123.92\\% \\ 
\emph{rest-ncs} & 84.2 & 15.9 & {\bf 0.00} & $< 0.001$ & 0.0 & 0.0 & 0.50 & 1.000 & 1000 & 1000 & 100.00\\% \\ 
\emph{rest-news} & 58.9 & 36.7 & {\bf 0.00} & $< 0.001$ & 1.8 & 0.3 & {\bf 0.07} & $< 0.001$ & 3218 & 4122 & 128.11\\% \\ 
\emph{rest-scs} & 62.6 & 11.7 & {\bf 0.00} & $< 0.001$ & 3.1 & 18.8 & {\bf 0.15} & $< 0.001$ & 1000 & 1000 & 100.00\\% \\ 
\emph{restcountries} & 68.2 & 15.0 & {\bf 0.00} & $< 0.001$ & 23.0 & 1.0 & {\bf 0.00} & $< 0.001$ & 1000 & 1000 & 100.00\\% \\ 
\emph{rpc-thrift-ncs} & 88.1 & 85.0 & {\bf 0.03} & $< 0.001$ & 10.0 & 10.0 & 0.50 & 1.000 & 5945 & 4915 & 82.67\\% \\ 
\emph{rpc-thrift-scs} & 71.7 & 74.5 & {\bf 0.83} & $< 0.001$ & 3.0 & 3.0 & 0.50 & 1.000 & 5512 & 5064 & 91.88\\% \\ 
\emph{scout-api} & 41.6 & 25.0 & {\bf 0.00} & $< 0.001$ & 43.7 & 8.7 & {\bf 0.00} & $< 0.001$ & 2666 & 7318 & 274.48\\% \\ 
\emph{session-service} & 64.7 & 57.7 & {\bf 0.28} & 0.003 & 18.7 & 10.8 & {\bf 0.04} & $< 0.001$ & 2467 & 1767 & 71.65\\% \\ 
\emph{timbuctoo} & 21.1 & 18.5 & {\bf 0.00} & $< 0.001$ & 24.7 & 8.6 & {\bf 0.00} & $< 0.001$ & 2795 & 8000 & 286.25\\% \\ 
\midrule 
Average  & 47.4 & 31.2 & 0.10 &  & 46.4 & 17.7 & 0.14 &  & 2980 & 4741 & 170.96\\% \\ 
Median  & 43.5 & 25.7 & 0.00 &  & 19.9 & 8.6 & 0.05 &  & 2795 & 4174 & 128.11\\% \\ 
\bottomrule 
\end{tabular} 

    %}
\end{table}

%\andrea{table width above must be fixed}

This section provides a foundational comparison of the MIO and RW algorithms.
Understanding the performance of these algorithms can establish a basis for investigating and interpreting the answers to the research questions. 
Table~\ref{table:mio_rw_comparison} presents comparison of the two algorithms. 
We follow the statistical guidelines from~\cite{Hitchhiker14}, reporting $p$-values of Mann-Whitney-Wilcoxon U tests and Vargha-Delaney standarized $\hat{A}_{12}$ effect sizes. 
Results are compared in terms of line coverage and detected faults.
To better understand the results, we also report the number of HTTP calls each algorithm execute, as each test case can have a different number of calls (from 1 to 10).
The computation cost and duration of a fitness function evaluation is directly related to the number of HTTP calls done in it.
In this case, RW takes longer, roughly 50\% more time.

When the line coverage results are examined, it is evident that the MIO algorithm achieves better outcomes in all cases except \emph{rpc-thrift-scs}, and all results are statistically significant. 
Regarding the number of detected faults, both algorithms exhibited the same performance in \emph{genome-nexus}, \emph{rest-ncs}, \emph{rpc-thrift-ncs}, and \emph{rpc-thrift-scs} problems. 
While MIO finds more faults in most problems, RW finds more faults than the MIO algorithm in the \emph{catwatch} and \emph{rest-scs} problems. 
The key result here is that according to the $\hat{A}_{12}$ metric, the MIO algorithm detects more faults in all problems. 
However, in some outlier runs, the number of faults detected by the RW algorithm is very high (e.g., 2700 in \emph{catwatch}, 202 in \emph{rest-scs} in one of the runs), and this increased the average.

As MIO provides better results than RW on average, we can use these results to support the choice of MIO to analyze the difficulty of branches compared to RW.

%-------------------------------------------------------------------------------------
\subsection{Selected Branches For Analysis}
\label{sub:branches}

\begin{table}[!t]
	\ifArxiv
	\centering
	\fi
	\footnotesize
	\caption{The number of branches reached and the number of branches that the algorithms reached but could not cover in any run.} \label{table:branch_info}
	\begin{tabular}{ l r r} 
\toprule 
SUT & \emph{\#Reached} & \emph{\#Never Covered} \\ 
\midrule 
\emph{catwatch} & 298 & 41 \\ 
\emph{cwa-verification} & 96 & 13 \\ 
\emph{genome-nexus} & 652 & 102 \\ 
\emph{gestaohospital-rest} & 50 & 9 \\ 
\emph{graphql-ncs} & 176 & 0 \\ 
\emph{graphql-scs} & 244 & 11 \\ 
\emph{languagetool} & 5083 & 1462 \\ 
\emph{market} & 30 & 3 \\ 
\emph{ocvn-rest} & 155 & 31 \\ 
\emph{patio-api} & 18 & 4 \\ 
\emph{pay-publicapi} & 14 & 1 \\ 
\emph{petclinic-graphql} & 22 & 1 \\ 
\emph{proxyprint} & 340 & 100 \\ 
\emph{reservations-api} & 16 & 5 \\ 
\emph{rest-ncs} & 168 & 0 \\ 
\emph{rest-news} & 104 & 21 \\ 
\emph{rest-scs} & 216 & 15 \\ 
\emph{restcountries} & 212 & 10 \\ 
\emph{rpc-thrift-ncs} & 168 & 0 \\ 
\emph{rpc-thrift-scs} & 218 & 7 \\ 
\emph{scout-api} & 279 & 41 \\ 
\emph{session-service} & 14 & 1 \\ 
\emph{timbuctoo} & 464 & 130 \\ 
\midrule 
Total & 9037 & 2008 \\ 
\bottomrule 
\end{tabular} 

\end{table}

Each branching statement/operation in the code has two possible outcomes: the constraint (i.e., a guarding condition that resolves to a boolean predicate) is satisfied and the ``then'' branch is followed, or it is not, and then the ``else'' branch is followed.
Unless an exception is thrown when evaluating the guarding condition, one of the two paths (i.e., ``then'' or ``else'') is necessarily taken.
In a fitness evaluation a branching condition might be evaluated $k>1$ times, e.g., if in a loop (e.g., \texttt{for} and \texttt{while} loops) or if in a function called more than once.
For calculating the fitness score, the highest value out of the $k$ branch distance evaluations is taken in \evo.

Note that, in SBST tools such as EvoSuite and \evo that work on JVM bytecode, branch coverage is computed at the bytecode level, and not at the code level.
For example, a single \texttt{if} statement in a Java class could results in many bytecode branch instructions, possibly one for each clause of the predicate in the guarding instruction.
In other words, the ``branch coverage'' criterion in EvoSuite (used in~\cite{albunian2020causes}) and in \evo (used in this study) is stronger, and more akin to the ``all-clauses'' coverage criterion.

To better clarify our analyses, let us introduce the concepts of \emph{reached} and \emph{covered} when discussing  branches.
If a test case generated during the search executes a branching statement, both the two resulting branches are marked as \emph{reached}.
If the predicate of that branch is resolved, then the ``then'' branch of the two will also be marked as \emph{covered}.
Otherwise, the ``else'' branch is marked as \emph{covered}.
If a branch is \emph{covered}, then it implies that it was also \emph{reached}.
The inverse is not necessarily true.
Furthermore, some branches may not have been ever reached during the search.
This happens for example in branches inside nested blocks, when the ``then'' branch is never entered and executed.
In this case, these branches are marked as \emph{never reached}.
A \emph{never reached} branch is implied to be \emph{never covered}.
However, a \emph{never covered} branch could had been \emph{reached}.

In our fitness landscape analyses, we only consider fitness scores for \emph{reached} branches.
All other \emph{never reached} branches would have the same score of 0 in all runs, so they would provide no useful information for our analyses.
Furthermore, we also excluded all \emph{reached} but \emph{never covered} branches.
These could be branches that are infeasible, or too hard to cover branches.
As their achieved data would not be enough to study their fitness landscape, those were excluded.
We also excluded branches that are covered when the SUT starts and boots up, before any test cases is executed.

Table~\ref{table:branch_info} provides details about the resulting analyzed branches of each SUT.
Although the algorithms reach up to \totalBranches branches in the study, \totalNeverCovered of these reached branches remain uncovered by any run or algorithm.
Therefore, \totalBranchesInFigures branches were used in the end for our fitness landscape analyses.

In the study we replicated, it is specified that 331 classes were employed for the analyses, although no details on the number of branches was provided in the text.
By analyzing the processed data in their replication study,\footnote{https://github.com/nasser-albunian/fitness-landscape-study}
it seems 3202 branches were used in the analyses (although we cannot be 100\% sure of this number).
If this is so, in our study we are looking at more than twice as many branches.

%As aforementioned,~\totalNeverCovered of the~\totalBranches branches were not covered at all during the study.
%Naturally, these branches do not contribute to the fitness landscape metrics and make it difficult to understand the behavior of other branches.
%Therefore, these branches were excluded from the analysis, and a total of~\totalBranchesInFigures branches were used.

For each branch,
the branch difficulty was assessed with the same approach as done in~\cite{albunian2020causes}, with four different labels.
A branch is classified as ``Easy'' if both the MIO and RW algorithms achieve at least a 50\% success rate (SR), i.e., if in both algorithms (total 60 runs) it was covered in at least 15 out of the 30 runs (per algorithm) of our experiments.
Conversely, if their success rates for both algorithms fall below 50\%, it is labeled ``Hard''.
If the success rate of the RW is under 50\% and the MIO's is above, it is labeled ``Search'';
if the situation is reversed, it is called ``RW''.

Note that the data for MIO is only used to assess the difficulty of the branches.
When computing the different metrics to study the characteristics of the fitness landscape, only the data for RW is used.

%--------------------------------------------------------------------------------
\subsection{Results for the RQ1}
\label{sub:rq1}

\begin{table}[t]
	\ifArxiv
	\centering
	\fi
    \footnotesize
    \caption{Mean six fitness landscape measurements for each SUT, gathered using the~\baseAlgorithmText algorithm.} \label{table:six-measure}
    \begin{tabular}{ l r r r r r r} 
\toprule 
\emph{SUT} & \emph{AC} & \emph{ND} & \emph{NV} & \emph{IC} & \emph{PIC}  & \emph{DBI} \\ 
\midrule 
catwatch & 0.836 & 0.774 & 0.002 & 0.046 & 0.002 & 0.017 \\ 
cwa-verification & 0.876 & 0.759 & 0.002 & 0.069 & 0.003 & 0.028 \\ 
genome-nexus & 0.884 & 0.874 & 0.002 & 0.034 & 0.002 & 0.012 \\ 
gestaohospital-rest & 0.804 & 0.932 & 0.002 & 0.015 & 0.001 & 0.005 \\ 
graphql-ncs & 0.736 & 0.785 & 0.006 & 0.116 & 0.012 & 0.058 \\ 
graphql-scs & 0.815 & 0.631 & 0.003 & 0.144 & 0.014 & 0.069 \\ 
languagetool & 0.984 & 0.994 & 0.001 & 0.002 & 0.000 & 0.001 \\ 
market & 0.933 & 0.984 & 0.001 & 0.006 & 0.000 & 0.002 \\ 
ocvn-rest & 0.714 & 0.889 & 0.001 & 0.026 & 0.003 & 0.009 \\ 
patio-api & 0.975 & 0.979 & 0.001 & 0.006 & 0.000 & 0.003 \\ 
pay-publicapi & 0.785 & 0.819 & 0.002 & 0.015 & 0.001 & 0.004 \\ 
petclinic-graphql & 0.890 & 0.718 & 0.001 & 0.083 & 0.005 & 0.035 \\ 
proxyprint & 0.977 & 0.980 & 0.001 & 0.004 & 0.000 & 0.002 \\ 
reservations-api & 0.874 & 0.848 & 0.001 & 0.039 & 0.002 & 0.016 \\ 
rest-ncs & 0.956 & 0.989 & 0.005 & 0.007 & 0.001 & 0.005 \\ 
rest-news & 0.796 & 0.812 & 0.002 & 0.053 & 0.007 & 0.023 \\ 
rest-scs & 0.953 & 0.927 & 0.004 & 0.033 & 0.005 & 0.018 \\ 
restcountries & 0.960 & 0.948 & 0.001 & 0.018 & 0.003 & 0.008 \\ 
rpc-thrift-ncs & 0.850 & 0.240 & 0.015 & 0.190 & 0.007 & 0.089 \\ 
rpc-thrift-scs & 0.847 & 0.511 & 0.005 & 0.089 & 0.003 & 0.036 \\ 
scout-api & 0.933 & 0.950 & 0.001 & 0.022 & 0.003 & 0.010 \\ 
session-service & 0.819 & 0.801 & 0.001 & 0.064 & 0.004 & 0.026 \\ 
timbuctoo & 0.687 & 0.711 & 0.002 & 0.082 & 0.009 & 0.034 \\ 
\midrule 
Average & 0.865 & 0.820 & 0.003 & 0.051 & 0.004 & 0.022 \\ 
Median & 0.874 & 0.848 & 0.002 & 0.034 & 0.003 & 0.016 \\ 
\bottomrule 
\end{tabular} 

\end{table}

\begin{figure}[t]
    \centering
    \includegraphics[width=.8\linewidth]{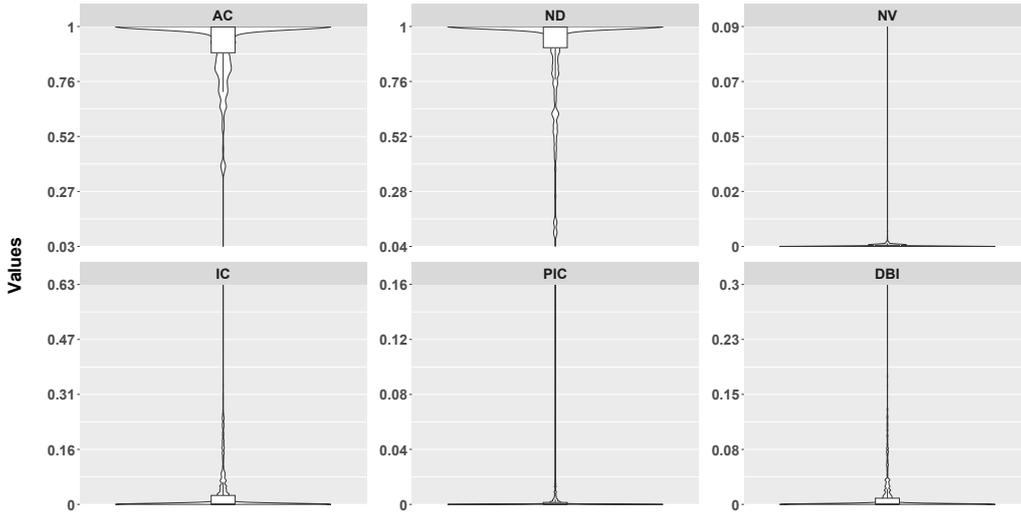}
    %\vspace{-2.5\baselineskip}
    \caption{Results of the six fitness landscape measures applied on the \totalBranchesInFigures branches. It is calculated by taking the average of 30 runs of~\baseAlgorithmText for each branch.}
    \label{fig:six_measures}
    %\vspace{-1.0\baselineskip}
\end{figure}

\begin{figure}[t]
    \centering
    \includegraphics[width=.8\linewidth]{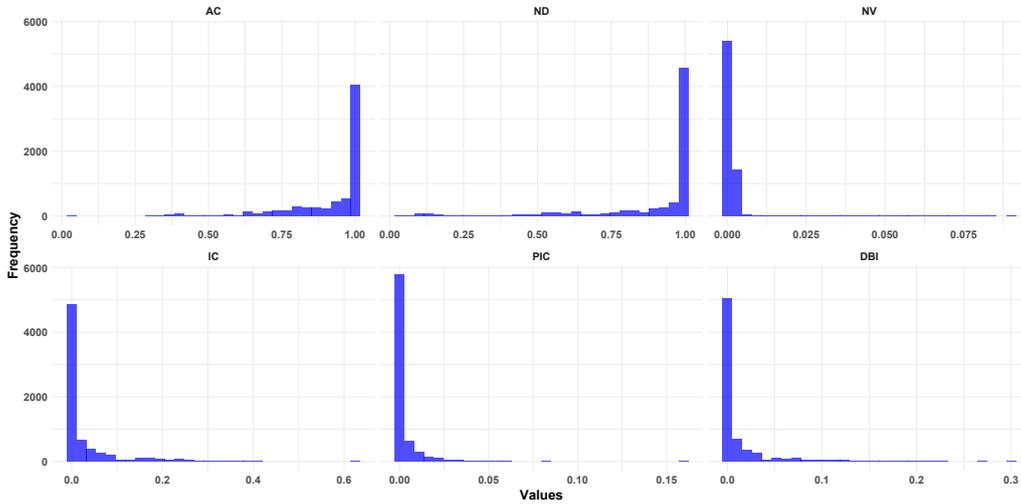}
    %\vspace{-2.5\baselineskip}
    \caption{Histogram of the six fitness landscape measures applied on the \totalBranchesInFigures branches. It is calculated by taking the average of 30 runs of~\baseAlgorithmText for each branch.}
    \label{fig:six_measures_histogram}
    %\vspace{-1.0\baselineskip}
\end{figure}

Table~\ref{table:six-measure} presents the results of the six fitness landscape measures applied to   the~\totalBranchesInFigures branches in our analyses.
The results are also visualized in Figure~\ref{fig:six_measures} and Figure~\ref{fig:six_measures_histogram}. 
The data given in both figures is the average metric value (i.e., the arithmetic mean) of each branch over 30 runs for RW.

According to Table~\ref{table:six-measure}, the lowest AC value is $0.687$, while the average is $0.865$.
These results indicate that the fitness landscape is highly correlated and smooth, with few peaks. 

When the ND metric is examined, the lowest value is $0.240$ only in the \emph{rpc-thrift-ncs} problem, while the others are greater than $0.5$.
This shows that the majority of the steps are neutral, and the fitness landscape generally consists of plateaus. 
According to the average values, $0.82$ of the steps consist of neutral steps. 

When the NV metric is examined, similar results are seen.
The highest value is $0.015$, which shows that the landscape generally consists of neutral areas. 

When the IC metric is examined, the highest value is $0.190$, while the average is $0.051$.
In general, the values are quite close to $0$. 
This shows that most branches have very low peaks and many flat areas. 

According to the PIC metric, the highest value is $0.014$, and the average is $0.004$.
This shows that branches generally consist of flat areas with very few slopes. 

When the DBI metric is examined, the highest value is $0.089$, and the average is $0.022$.
As for IC and PIC, the low values for the DBI metric do not suggest that the landscape is rugged (recall Table~\ref{tab:metrics}).

%Web APIs are different from OO software from the point of view of testing. 
%Operations on an object will all be on the same state by construction. 
%For example, x.foo(); x.bar() are both on the same object x.
%This is more difficult to achieve in Web APIs, as they might deal with different states when randomizing actions (e.g., /x/\{id\} would need some concrete id to refer to the same state). 
%Therefore, when a request is made to an endpoint, a mutation (one step forward) applied to a branch while running may cause the corresponding branch value to be $0$. 
%Consequently, the DBI value is generally close to $0$, indicating the fitness landscape is rugged. \omur{We need to investigate the above comment in detail.}

When analyzing Figures~\ref{fig:six_measures} and~\ref{fig:six_measures_histogram}, we see that although the branches' values for AC and ND metrics fall within the range of $[0, 1]$, the majority have a value of $0$. 
Similarly, for the NV metric, which ranges from $[0, 0.09]$, most branches again only show a value of $0$. 
For the IC metric, values are within $[0, 0.63]$, but again, the majority of branches have a value of $0$. 
The PIC metric ranges from $[0, 0.16]$, yet most branches still have a value of $0$. 
Lastly, regarding the DBI metric, although values vary from $[0, 0.3]$, most branches again have a value of $0$.

\begin{result}
{\bf RQ1:} Neutrality prevails across the fitness landscape, where most branches appear relatively smooth. 
Only very few branches seem rugged.
\end{result}

%-----------------------------------------------------------------------------------------------------
\subsection{Results for the RQ2}

\begin{figure}[t]
    \centering
    \includegraphics[width=.8\linewidth]{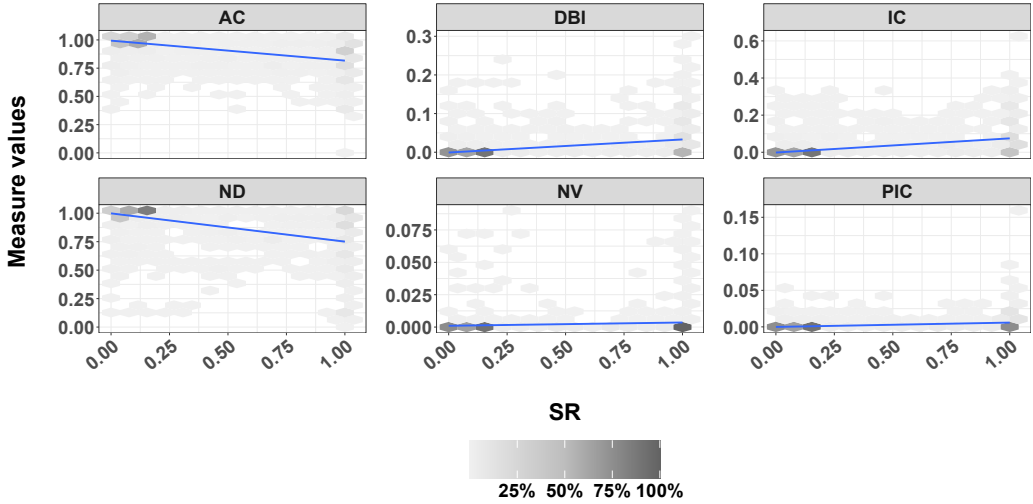}
    %\vspace{-2.5\baselineskip}
    \caption{The Spearman correlation of Success Rate (SR) with each of the six measures for all of the \totalBranchesInFigures branches.
    While the Y-axis presents the value of each metric, the X-axis shows the SR.
    Each hexagon represents a group of runs for different branches, with the density of the hexagons increasing as the number of runs within a hexagon grows.
    %The correlation coefficient of SR and AC is~\correlationAC, ND is~\correlationND, NV is~\correlationNV, IC is~\correlationIC, PIC is~\correlationPIC, and DBI is~\correlationDBI.
    }
    \label{fig:correlation}
    %\vspace{-1.0\baselineskip}
\end{figure}

Spearman correlation values between success rates (SR) and each metric were calculated. Spearman's correlation is a statistical measure that quantifies the monotonic relationship between two variables. 
A monotonic relationship indicates that as one variable increases, the other tends to increase (positive correlation) or decrease (negative correlation).
According to these calculations, the correlation coefficient of SR and AC is~\correlationAC, ND is~\correlationND, NV is~\correlationNV, IC is~\correlationIC, PIC is~\correlationPIC, and DBI is~\correlationDBI.
Figure~\ref{fig:correlation} examines the effect of fitness landscape features.

When Figure~\ref{fig:correlation} is examined, we can see that the SR increases as the AC metric decreases.
In other words, it can be said that test generation becomes more difficult in cases where fitness values have a high correlation.
When the ND metric is examined, the success rate decreases as the neutrality distance increases. 
When examining the NV metric, the diversity of fitness values makes it easier to reach a solution. 
The ability to generate a solution increases as the NV value increases. 
Analyzing the IC reveals that greater information content enhances the SR.
Examining the PIC shows that an expanded landscape modality also boosts the SR.
When looking at the DBI metric, a higher variety of flat areas within the landscape leads to increased success.
As a result, while there is a negative relationship between the AC and ND metrics and SR, there is a positive relationship between the NV, IC, PIC, and DBI metrics.

Taking into account the interpretations of these metrics form Table~\ref{tab:metrics},
and taking into account the obtained Spearman correlation values,
an increase in neutrality (i.e., high ND and low NV) leads to worse results (i.e., lower SR).
However, more ruggedness  (i.e., AC closer to 0, and higher values for IC, PIC and DBI) seems to lead to better results.

\begin{result}
{\bf RQ2:} %Fitness landscape properties like AC and ND make system-level test generation harder. 
%Meanwhile, greater NV, IC, PIC, and DBI enhance it, underscoring the need for balanced landscape design.
Increasing the correlation between solutions (AC) or the number of neutral steps (ND) makes system-level test generation more challenging. 
Conversely, increasing the number of distinctive regions (NV), the number of peaks in the landscape (IC), the modality (PIC), or the diversity of neutral areas (DBI) makes system-level test generation easier.
In our analyses,  while neutrality leads to more challenges in the search, an increase in ruggedness seems on the other hand to make it is easier.
\end{result}

%-----------------------------------------------------------------------------------------------------
\subsection{Results for the RQ3}

\begin{table}[t]
	\ifArxiv
	\centering
	\fi
    \footnotesize
    \caption{Six fitness landscape measurements for each group, gathered using the~\baseAlgorithmText algorithm.
    $NB$ represent the size of each group (i.e., number of branches in each group), considering a total of \totalBranchesInFigures branches.} \label{table:six-measure-group}
    \begin{tabular}{ l r r r r r r r} 
\toprule 
\emph{GROUP} & \emph{NB}  & \emph{AC} & \emph{ND} & \emph{NV} & \emph{IC} & \emph{PIC}  & \emph{DBI} \\ 
\midrule 
Easy & 1429 & 0.763 & 0.651 & 0.004 & 0.099 & 0.007 & 0.043 \\ 
Search & 1047 & 0.908 & 0.923 & 0.003 & 0.029 & 0.003 & 0.014 \\ 
Hard & 4437 & 0.981 & 0.980 & 0.001 & 0.005 & 0.000 & 0.002 \\ 
RW & 116 & 0.787 & 0.639 & 0.002 & 0.090 & 0.007 & 0.037 \\ 
\bottomrule 
\end{tabular} 

\end{table}

In the previous sections, we have observed how features of the fitness landscape can influence search outcomes.
In this section, we will investigate which features of the code affect the search process. 
To detail this, we categorize each branch based on the success of the search algorithm (MIO) compared to the random walk (RW).

\begin{figure}[t]
    \centering
    \includegraphics[width=.8\linewidth]{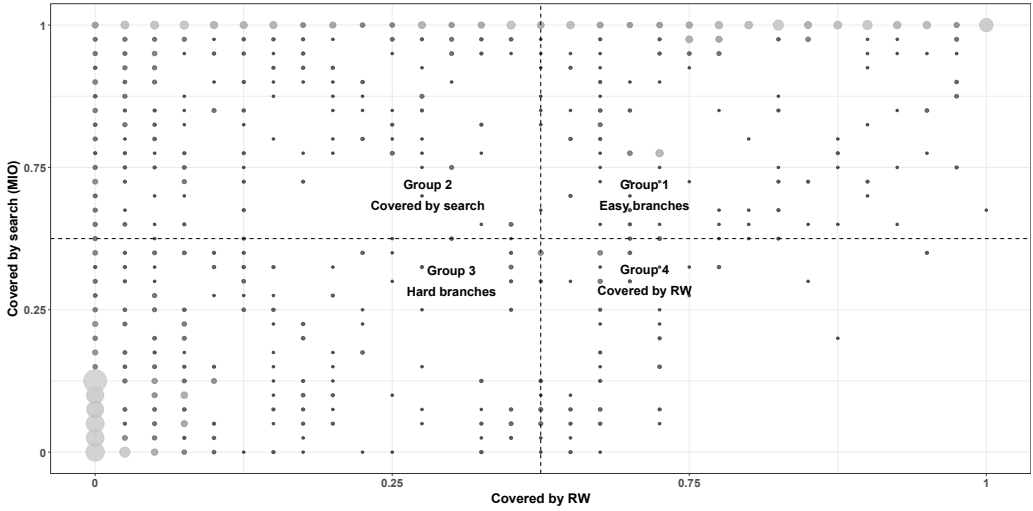}
    %\vspace{-2.5\baselineskip}
    \caption{Four groups of the branches based on their coverage by \compareAlgorithmText and \baseAlgorithmText where a large bubble size indicates a high number of branches}
    \label{fig:four_groups}
    %\vspace{-1.0\baselineskip}
\end{figure}

\begin{figure}[t]
    \centering
    \includegraphics[width=.8\linewidth]{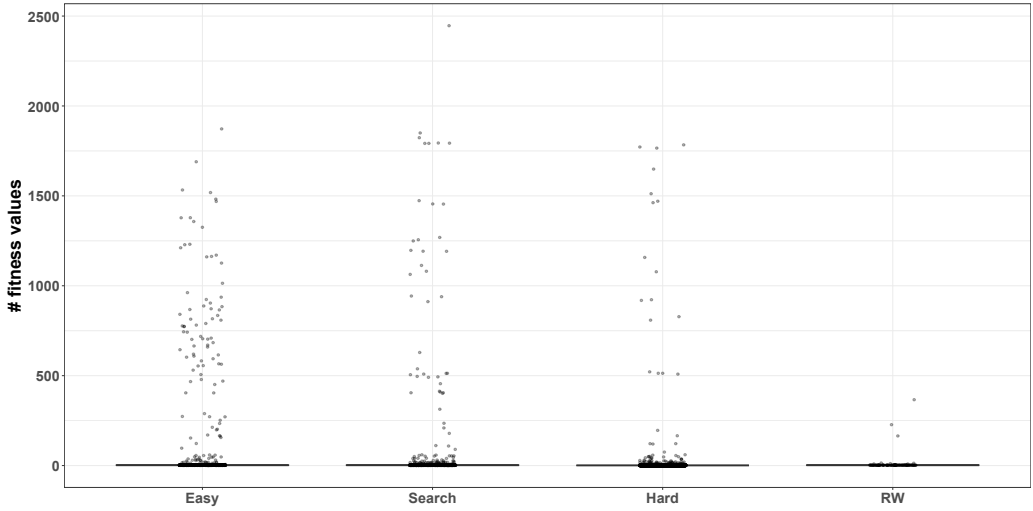}
    %\vspace{-2.5\baselineskip}
    \caption{Number of discrete fitness values obtained by the random walk for each branch in the four groups}
    \label{fig:discrete_fitness}
    %\vspace{-1.0\baselineskip}
\end{figure}

\begin{figure}[t]
    \centering
    \includegraphics[width=.8\linewidth]{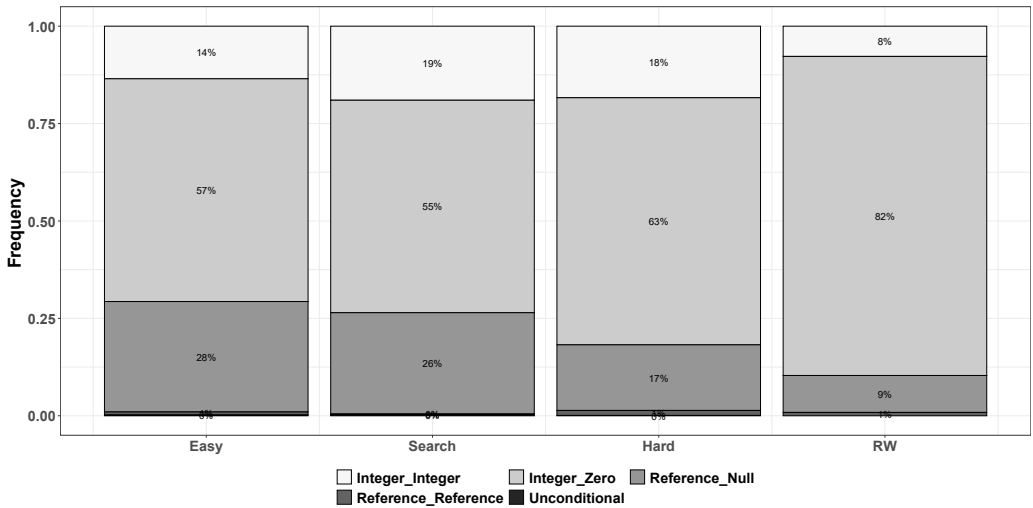}
    %\vspace{-2.5\baselineskip}
    \caption{Classifications of the branch types in the four groups}
    \label{fig:classification}
    %\vspace{-1.0\baselineskip}
\end{figure}

As discussed in Section~\ref{sub:branches},
the ``Easy'' group contains branches where both the RW and MIO algorithms have success rates greater than 50\%.
The ``Hard'' group includes branches with success rates below 50\% for both algorithms.
The ``Search'' group is characterized by a success rate above 50\% for the MIO algorithm and below 50\% for the RW algorithm, while the ``RW'' group is the opposite.
Table~\ref{table:six-measure-group} and Figures~\ref{fig:four_groups},~\ref{fig:discrete_fitness}, and~\ref{fig:classification} are created according to these distinctions.

When the Table~\ref{table:six-measure-group} is examined, we can see that in the Search group both AC ($0.908$) and ND ($0.923$) values are relatively high, indicating that the landscape is smoother and contains a greater density of neutral areas.
%The NV value ($0.003$) is average \andrea{what does this mean?} compared to other groups, suggesting that the diversity of fitness values is moderate.
The NV value ($0.003$) is higher than that of the RW and Hard groups but lower than that of the Easy group.
According to the IC metric ($0.029$), the landscape's ruggedness, measured by the number of peaks, is lower than that of the Easy and RW groups but higher than that of the Hard group. 
The PIC metric ($0.003$) indicates that the modality is greater than that of the Hard group, while the DBI metric ($0.014$) shows that the diversity of flat areas is more than in the Hard group but less than in the Easy and RW groups. 
When examining the Hard group, which contains most of the branches, it is observed that the landscape is smooth ($AC = 0.981$), with a very high level of neutral areas ($ND = 0.980$). 
However, the NV is relatively low at $0.001$, suggesting that it typically has a single fitness value that cannot be improved. 
The number of peaks in this landscape ($IC = 0.005$) and the modality ($PIC = 0.000$) are both low. 
Additionally, the diversity of flat areas is quite limited ($DBI = 0.002$). 
Finally, as anticipated, there are only few branches in the RW group.
Generally, the metrics for this group are similar to those of the Easy group, with the NV metric ($0.002$) being the main difference. 
While the diversity of neutral areas in RW group is lower than that in the Easy group, it remains higher than in the other groups. However, the diversity of fitness values is lower, which may have caused challenges for the search algorithms in covering it.
As a result, these branches may have been more effectively addressed by the RW group, which generates more random solutions.
%Since the fitness diversity in the RW group is lower than in the Easy group, it may have been challenging for the search algorithm to cover branches 

Upon examining Figure~\ref{fig:four_groups}, it becomes clear that most results fall within the Hard class (bottom left). 
In this category, most branches remain uncovered by either algorithm in most of the runs.
It is important to note that branches not covered by any algorithm across their 30 runs were excluded from the evaluation, i.e., the data for Hard branches still include at least one successful run out of 60 runs.
Another significant group comprises the Easy branches covered by both algorithms (top right).

Figure~\ref{fig:discrete_fitness} gives the number of distinct fitness values of branches in different groups. 
The number of different fitness values is generally close to 1 in all groups. 
It should be noted here that a fitness value is in the range of $[0,1]$ and is based only on branch distance. 
This restricts the diversity of fitness values considerably. 
A slight slope in the landscape allows for different fitness values. 
There are a high number of different fitness values for only a limited number of branches. 
This shows that the landscape's slope is quite low and generally consists of flat areas. 
It is seen that the highest number of different fitness values are in the Easy group and the lowest in the RW group.
Unsurprisingly, the fitness diversity in the RW group is low because a search algorithm like MIO directs the search based on gradient information and tries to find the appropriate solution.
Since RW is based on random mutations, it does not need gradient information or direction mechanisms like fitness function. 
Therefore, the diversity created by mutations can potentially overcome the limitations of gradient-based or search-based optimization methods.

All groups, except for RW, show similar numbers of discrete fitness function values. 
This suggests that the ruggedness of the groups does not differ significantly. 
Therefore, a more detailed analysis of the branches is necessary to evaluate their performance differences.
Figure~\ref{fig:classification} provides this detailed examination, where each branch is categorized based on their bytecode type, as specified in more details in reference \cite{shamshiri2018random}.

One thing to consider is that \evo has reported an ``Unconditional'' branch type that includes the ``goto'' statement.
This is usually related to \texttt{continue} and \texttt{break} statements in loops.
As this is not technically considered as a branch, it can be considered as a fault in \evo, which will be fixed in a future release.
Therefore, although for consistency these ``branches'' are still listed in the following tables, they are not discussed in any further detail.
As only $10$ cases fall into this category, they do not significantly impact any of our results.

%Additionally, a new category called the ``Unconditional'' branch type\andrea{do we have branch distances for this?}, which includes the ``goto'' statement, has been introduced.
%It should be noted that at the bytecode level, this is a goto statement \andrea{what you mean?}.
%This can be done with \texttt{continue} or \texttt{break} statements.
%In the cases of the \emph{languagetool} and \emph{rest-news} problems, only $10$ branches fall into this category \andrea{what about others?}.

Upon examining Figure~\ref{fig:classification}, it becomes evident that most groups consist of branches of the ``Integer\_Zero'' type. 
This branch type primarily corresponds to the \emph{if(x)} boolean expression translated by the Java compiler, contributing to the presence of plateaus in the fitness landscape~(\cite{Mcm04}), i.e., a typical source of \emph{flag problem}~(\cite{HHH02}).
In the Search group, the number of ``Integer\_Integer'' type branches is slightly higher than in the other groups. 
Notably, this branch type is the only one that creates a clear gradient in the fitness landscape.

Conversely, in the Easy group, there is a higher prevalence of ``Reference\_Null'' type branches compared to the other groups.
Although the ``Reference\_Null'' and ``Reference\_Reference'' type benchmarks exhibit lower discrete fitness values, they are anticipated to be complicated. 
Interestingly, most branches of the RW are of the ``Integer\_Zero'' type, which may indicate the presence of complex predicates that are challenging for an optimization algorithm to resolve due to a lack of fitness gradient.

\begin{landscape}
    \ifSpringer
	\setlength{\topmargin}{3cm}
	\fi
    \begin{table*} \tiny
    \caption{Comparison of the performance of the \baseAlgorithmText and \compareAlgorithmText algorithms in different branch types. NB represents the number of branches, and SR indicates the success rate. The ``Integer\_Integer'' column illustrates branches where two integer values are compared, ``Integer\_Zero'' shows comparisons between an integer value and 0, ``Reference\_Reference'' signifies the comparison of two reference values, ``Reference\_Null'' compares a reference value to a null value, and the final category covers the ``Unconditional'' which is ``goto'' statement.} \label{table:branch_types}
    \begin{tabular}{l|rrrr|rrrr|rrrr|rrrr|rrrr}
\toprule
& \multicolumn{4}{c|}{Integer\_Integer} &\multicolumn{4}{c|}{Integer\_Zero} &\multicolumn{4}{c|}{Reference\_Reference} &\multicolumn{4}{c|}{Reference\_Null} &\multicolumn{4}{c}{Unconditional} \\ \hline
SUTS & NB & $SR_{RW}$ & $SR_{MIO}$ & $\hat{A}_{12}$ & NB & $SR_{RW}$ & $SR_{MIO}$ & $\hat{A}_{12}$ & NB & $SR_{RW}$ & $SR_{MIO}$ & $\hat{A}_{12}$ & NB & $SR_{RW}$ & $SR_{MIO}$ & $\hat{A}_{12}$ & NB & $SR_{RW}$ & $SR_{MIO}$ & $\hat{A}_{12}$\\ \hline 
\emph{catwatch} & 14  &  0.32  &  0.47  &  0.70  &  151  &  0.39  &  0.51  &  \textbf{0.68}  &  1  &  0.10  &  0.30  &  1.00  &  91  &  0.43  &  0.63  &  \textbf{0.72}  &  0  &  \emph{NA}  &  \emph{NA}  &  \emph{NA}  \\ \hline 
\emph{cwa-verification} & 6  &  0.28  &  0.47  &  0.67  &  46  &  0.43  &  0.55  &  0.62  &  6  &  0.59  &  0.65  &  0.44  &  25  &  0.51  &  0.61  &  0.59  &  0  &  \emph{NA}  &  \emph{NA}  &  \emph{NA}  \\ \hline 
\emph{genome-nexus} & 78  &  0.42  &  0.61  &  \textbf{0.66}  &  262  &  0.47  &  0.62  &  \textbf{0.64}  &  0  &  \emph{NA}  &  \emph{NA}  &  \emph{NA}  &  210  &  0.56  &  0.77  &  \textbf{0.75}  &  0  &  \emph{NA}  &  \emph{NA}  &  \emph{NA}  \\ \hline 
\emph{gestaohospital-rest} & 6  &  0.33  &  0.80  &  0.85  &  25  &  0.27  &  0.59  &  \textbf{0.77}  &  0  &  \emph{NA}  &  \emph{NA}  &  \emph{NA}  &  10  &  0.37  &  0.72  &  \textbf{0.79}  &  0  &  \emph{NA}  &  \emph{NA}  &  \emph{NA}  \\ \hline 
\emph{graphql-ncs} & 72  &  0.41  &  0.79  &  \textbf{0.83}  &  104  &  0.45  &  0.85  &  \textbf{0.87}  &  0  &  \emph{NA}  &  \emph{NA}  &  \emph{NA}  &  0  &  \emph{NA}  &  \emph{NA}  &  \emph{NA}  &  0  &  \emph{NA}  &  \emph{NA}  &  \emph{NA}  \\ \hline 
\emph{graphql-scs} & 50  &  0.33  &  0.60  &  \textbf{0.70}  &  181  &  0.49  &  0.48  &  \textbf{0.43}  &  1  &  0.40  &  1.00  &  1.00  &  1  &  0.60  &  1.00  &  1.00  &  0  &  \emph{NA}  &  \emph{NA}  &  \emph{NA}  \\ \hline 
\emph{languagetool} & 643  &  0.01  &  0.11  &  \textbf{0.88}  &  2278  &  0.03  &  0.14  &  \textbf{0.89}  &  55  &  0.03  &  0.15  &  \textbf{0.96}  &  641  &  0.08  &  0.22  &  \textbf{0.89}  &  4  &  0.00  &  0.15  &  \textbf{1.00}  \\ \hline 
\emph{market} & 0  &  \emph{NA}  &  \emph{NA}  &  \emph{NA}  &  20  &  0.33  &  0.44  &  0.61  &  0  &  \emph{NA}  &  \emph{NA}  &  \emph{NA}  &  7  &  0.39  &  0.53  &  0.63  &  0  &  \emph{NA}  &  \emph{NA}  &  \emph{NA}  \\ \hline 
\emph{ocvn-rest} & 11  &  0.27  &  0.56  &  0.69  &  52  &  0.33  &  0.74  &  \textbf{0.79}  &  0  &  \emph{NA}  &  \emph{NA}  &  \emph{NA}  &  61  &  0.48  &  0.91  &  \textbf{0.90}  &  0  &  \emph{NA}  &  \emph{NA}  &  \emph{NA}  \\ \hline 
\emph{patio-api} & 0  &  \emph{NA}  &  \emph{NA}  &  \emph{NA}  &  14  &  0.12  &  0.48  &  \textbf{0.83}  &  0  &  \emph{NA}  &  \emph{NA}  &  \emph{NA}  &  0  &  \emph{NA}  &  \emph{NA}  &  \emph{NA}  &  0  &  \emph{NA}  &  \emph{NA}  &  \emph{NA}  \\ \hline 
\emph{pay-publicapi} & 2  &  0.50  &  0.50  &  0.50  &  9  &  0.57  &  0.67  &  0.54  &  0  &  \emph{NA}  &  \emph{NA}  &  \emph{NA}  &  2  &  0.57  &  1.00  &  0.75  &  0  &  \emph{NA}  &  \emph{NA}  &  \emph{NA}  \\ \hline 
\emph{petclinic-graphql} & 0  &  \emph{NA}  &  \emph{NA}  &  \emph{NA}  &  4  &  0.39  &  1.00  &  \textbf{1.00}  &  0  &  \emph{NA}  &  \emph{NA}  &  \emph{NA}  &  17  &  0.38  &  0.93  &  \textbf{0.94}  &  0  &  \emph{NA}  &  \emph{NA}  &  \emph{NA}  \\ \hline 
\emph{proxyprint} & 14  &  0.07  &  0.14  &  \textbf{0.23}  &  105  &  0.07  &  0.09  &  \textbf{0.18}  &  1  &  0.00  &  0.03  &  1.00  &  120  &  0.11  &  0.53  &  \textbf{0.65}  &  0  &  \emph{NA}  &  \emph{NA}  &  \emph{NA}  \\ \hline 
\emph{reservations-api} & 0  &  \emph{NA}  &  \emph{NA}  &  \emph{NA}  &  8  &  0.65  &  0.70  &  0.62  &  0  &  \emph{NA}  &  \emph{NA}  &  \emph{NA}  &  3  &  0.63  &  1.00  &  0.83  &  0  &  \emph{NA}  &  \emph{NA}  &  \emph{NA}  \\ \hline 
\emph{rest-ncs} & 72  &  0.09  &  0.79  &  \textbf{0.95}  &  96  &  0.09  &  0.81  &  \textbf{0.92}  &  0  &  \emph{NA}  &  \emph{NA}  &  \emph{NA}  &  0  &  \emph{NA}  &  \emph{NA}  &  \emph{NA}  &  0  &  \emph{NA}  &  \emph{NA}  &  \emph{NA}  \\ \hline 
\emph{rest-news} & 0  &  \emph{NA}  &  \emph{NA}  &  \emph{NA}  &  29  &  0.40  &  0.82  &  \textbf{0.85}  &  0  &  \emph{NA}  &  \emph{NA}  &  \emph{NA}  &  48  &  0.37  &  0.80  &  \textbf{0.85}  &  6  &  0.57  &  0.83  &  0.76  \\ \hline 
\emph{rest-scs} & 46  &  0.06  &  0.57  &  \textbf{0.78}  &  154  &  0.08  &  0.46  &  \textbf{0.66}  &  1  &  0.10  &  1.00  &  1.00  &  0  &  \emph{NA}  &  \emph{NA}  &  \emph{NA}  &  0  &  \emph{NA}  &  \emph{NA}  &  \emph{NA}  \\ \hline 
\emph{restcountries} & 30  &  0.07  &  0.94  &  \textbf{1.00}  &  155  &  0.07  &  0.89  &  \textbf{0.99}  &  0  &  \emph{NA}  &  \emph{NA}  &  \emph{NA}  &  17  &  0.16  &  0.91  &  \textbf{1.00}  &  0  &  \emph{NA}  &  \emph{NA}  &  \emph{NA}  \\ \hline 
\emph{rpc-thrift-ncs} & 72  &  0.79  &  0.84  &  0.52  &  96  &  0.78  &  0.86  &  0.54  &  0  &  \emph{NA}  &  \emph{NA}  &  \emph{NA}  &  0  &  \emph{NA}  &  \emph{NA}  &  \emph{NA}  &  0  &  \emph{NA}  &  \emph{NA}  &  \emph{NA}  \\ \hline 
\emph{rpc-thrift-scs} & 46  &  0.70  &  0.70  &  0.52  &  164  &  0.75  &  0.67  &  0.45  &  1  &  1.00  &  1.00  &  \emph{NA}  &  0  &  \emph{NA}  &  \emph{NA}  &  \emph{NA}  &  0  &  \emph{NA}  &  \emph{NA}  &  \emph{NA}  \\ \hline 
\emph{scout-api} & 39  &  0.11  &  0.45  &  \textbf{0.75}  &  117  &  0.12  &  0.54  &  \textbf{0.83}  &  1  &  0.00  &  0.13  &  1.00  &  81  &  0.18  &  0.68  &  \textbf{0.88}  &  0  &  \emph{NA}  &  \emph{NA}  &  \emph{NA}  \\ \hline 
\emph{session-service} & 0  &  \emph{NA}  &  \emph{NA}  &  \emph{NA}  &  8  &  0.32  &  0.40  &  0.52  &  1  &  0.97  &  1.00  &  1.00  &  4  &  0.29  &  0.50  &  0.50  &  0  &  \emph{NA}  &  \emph{NA}  &  \emph{NA}  \\ \hline 
\emph{timbuctoo} & 15  &  0.54  &  0.66  &  0.63  &  219  &  0.54  &  0.78  &  \textbf{0.74}  &  3  &  0.78  &  0.99  &  1.00  &  97  &  0.70  &  0.96  &  \textbf{0.86}  &  0  &  \emph{NA}  &  \emph{NA}  &  \emph{NA}  \\ \hline 
\end{tabular} 

    \end{table*}
\end{landscape}

Table~\ref{table:branch_types} compares the performance of the RW and MIO algorithms in different branch types. 
In the table, NB represents number of branches, while SR means the success rate.
The average success rates of the branches are presented in the table. 
Statistical analyses were performed using the average success rate value of each branch. 
In cases where the Mann-Whitney-Wilcoxon test result is $p<0.05$, the $\hat{A}_{12}$ value is shown in bold.

When ``Integer\_Integer'' type branches are examined, it is observed that the success rate of the MIO algorithm is significantly better. 
The most interesting of these results is \emph{proxyprint} in terms of $\hat{A}_{12}$. 
There are $14$ branches in total, and in $2$ of these branches, the MIO algorithm has a 100\% success rate, while in the remaining $12$ branches, it has a 0\% success rate. 
RW could not achieve a 100\% success rate in any branch, but its success rate is higher than $0$ in $12$ branches (usually, it is successful in $1$ out of $30$ runs). 
Therefore, MIO is seen to be more successful on average, while RW shows better performance according to $\hat{A}_{12}$ values.
When ``Integer\_Zero'' branches, which generally cause boolean flags, are examined, it is noted that most branches are ``Integer\_Zero'' type branches. 
This may indicate that the landscape has boolean flags in general. 
Except for \emph{graphql-scs}, \emph{proxyprint}, and \emph{rpc-thrift-scs}, it is observed that the MIO algorithm has a higher success rate. 
When we examine ``Reference\_Reference'' type branches, it is noted that very few branches are of this reference type.
Statistically, it is the branch type where the MIO and RW algorithms can come closest to each other. 
When we examine ``Reference\_Null'' type branches, it is seen that MIO is generally more successful.

%The number of branches in the Unconditional type is very low ($10$ in total).
%These statements can be realized by converting break and continue statements in a code snippet into bytecode. 
%When we examine it in general, it is seen that the search algorithm is more successful. 
%Still, it is hard to make a detailed analysis because of their very low numbers and the rare use of the goto \andrea{we need to discuss this in our meeting} statement as bytecode.

\begin{table}[t]
	\ifArxiv
	\centering
	\fi
	\tiny
	\caption{Branch types of the reached but never covered branches, for each SUT.} \label{table:never-covered-branch-types}
	\begin{tabular}{l|rrrrr} 
\toprule 
SUTS & Integer\_Integer & Integer\_Zero & Reference\_Reference & Reference\_Null & Unconditional \\ 
\midrule 
\emph{catwatch} & 0 & 11 & 1 & 29 & 0 \\ 
\emph{cwa-verification} & 0 & 4 & 0 & 9 & 0 \\ 
\emph{genome-nexus} & 6 & 38 & 0 & 58 & 0 \\ 
\emph{gestaohospital-rest} & 0 & 3 & 0 & 6 & 0 \\ 
\emph{graphql-scs} & 0 & 9 & 1 & 1 & 0 \\ 
\emph{languagetool} & 185 & 818 & 41 & 412 & 6 \\ 
\emph{market} & 0 & 0 & 0 & 3 & 0 \\ 
\emph{ocvn-rest} & 1 & 11 & 0 & 19 & 0 \\ 
\emph{patio-api} & 0 & 4 & 0 & 0 & 0 \\ 
\emph{pay-publicapi} & 0 & 1 & 0 & 0 & 0 \\ 
\emph{petclinic-graphql} & 0 & 0 & 0 & 1 & 0 \\ 
\emph{proxyprint} & 0 & 27 & 3 & 70 & 0 \\ 
\emph{reservations-api} & 0 & 2 & 0 & 3 & 0 \\ 
\emph{rest-news} & 0 & 1 & 0 & 20 & 0 \\ 
\emph{rest-scs} & 0 & 14 & 1 & 0 & 0 \\ 
\emph{restcountries} & 0 & 7 & 0 & 3 & 0 \\ 
\emph{rpc-thrift-scs} & 2 & 4 & 1 & 0 & 0 \\ 
\emph{scout-api} & 0 & 11 & 1 & 29 & 0 \\ 
\emph{session-service} & 0 & 0 & 1 & 0 & 0 \\ 
\emph{timbuctoo} & 1 & 67 & 3 & 59 & 0 \\ 
\midrule 
Total & 195 & 1032 & 53 & 722 & 6 \\ 
\bottomrule 
\end{tabular} 

\end{table}

Finally, Table~\ref{table:never-covered-branch-types} provides information about the branches that are reached but never covered, grouped according to their branch type.
According to this table, the most uncovered branch belongs to the \emph{languagetool} SUT.
Here, the most frequently encountered branch type is the ``Integer\_Zero'' type. 
Algorithms struggle to find solutions due to plateaus created by boolean flags and the complexity of predicates. 
Second in line are the ``Reference\_Null'' type branches. 
These branches are usually caused by null checks that prevent application crashes caused by null pointer exceptions or null check transformations in null safety languages such as Kotlin. 
These scenarios occur when a \texttt{null} value is not anticipated within the program's normal flow.

For instance, in the \emph{graphql-scs} problem, the code expression \texttt{"fileparts = file.split(".".toRegex()).toTypedArray()"} has been translated into the statements depicted in Figure~\ref{code:translatedKotlinGraphQlScs}, which has subsequently been converted to bytecode shown in Figure~\ref{code:translatedKotlinGraphQlScsByteCode}.
Covering the branch in this expression proves to be quite challenging.
Furthermore, branches that could be covered in unit testing (e.g., by passing a \texttt{null} as input parameter) might be \emph{infeasible} in system testing, as any input to the function has to go through the user input interfaces (e.g., HTTP calls in the case of REST APIs), which might do some input sanitization and filtering.

\begin{figure}
    \begin{lstlisting}[language=java]
CharSequence var7 = (CharSequence)file;
Regex var8 = new Regex(".");
byte var9 = 0;
Collection $this$toTypedArray$iv = (Collection)var8.split(var7, var9);
int $i$f$toTypedArray = false;
Object[] var10000 = $this$toTypedArray$iv.toArray(new String[0]);
if (var10000 == null) {
 throw new NullPointerException("null cannot be cast to non-null type kotlin.Array<T of kotlin.collections.ArraysKt__ArraysJVMKt.toTypedArray>");
} else {
 fileparts = (String[])var10000;
	...
}
    \end{lstlisting}\caption{Code snipped from FileSuffix class in graphql-scs problem.}\label{code:translatedKotlinGraphQlScs}
\end{figure}

\begin{figure}
    \begin{lstlisting}[language=java]
 aload 9
 iconst_0
 anewarray java/lang/String
 invokeinterface java/util/Collection.toArray([Ljava/lang/Object;)[Ljava/lang/Object;
 dup
 ifnonnull L12
 pop
 new java/lang/NullPointerException
 dup
 ldc "null cannot be cast to non-null type kotlin.Array<T of kotlin.collections.ArraysKt__ArraysJVMKt.toTypedArray>" (java.lang.String)
 invokespecial java/lang/NullPointerException.<init>(Ljava/lang/String;)V
 athrow
\end{lstlisting}\caption{Bytecodes of the FileSuffix class given in Figure~\ref{code:translatedKotlinGraphQlScs}}\label{code:translatedKotlinGraphQlScsByteCode}
\end{figure}

\begin{result}
{\bf RQ3:}
Most plateaus in the fitness landscape result from boolean flags and null checks used to prevent crashes in the code.
This creates branches that are difficult to cover, or simply infeasible in system testing.
\end{result}

%----------------------------------------------------------------------------------------------
\subsection{Results for the RQ4}

\begin{table}[t]
	\ifArxiv
	\centering
	\fi
    \footnotesize
    \caption{Six fitness landscape measurements obtained for each group in the baseline study~\cite{albunian2020causes}.} \label{table:six-measure-group-baseline}
    \begin{tabular}{ l r r r r r r r} 
\toprule 
\emph{GROUP} & \emph{AC} & \emph{ND} & \emph{NV} & \emph{IC} & \emph{PIC}  & \emph{DBI} \\
\midrule 
Easy & 0.652 & 0.114 & 0.009 & 0.401 & 0.092 & 0.872 \\
Search & 0.829 & 0.129 & 0.005 & 0.125 & 0.058 & 0.904 \\
Hard & 0.898 & 0.516 & 0.002 & 0.076 & 0.028 & 0.960 \\
RW & 0.852 & 0.258 & 0.004 & 0.098 & 0.040 & 0.928 \\
\bottomrule 
\end{tabular} 

\end{table}

Table~\ref{table:six-measure-group-baseline} presents the results of six different fitness landscape metrics published in the baseline study~\cite{albunian2020causes}.
Recall that the values reported by the various groups in this study are outlined in Table~\ref{table:six-measure-group}.
Examining both tables reveals a parallelism in AC values, except for the RW algorithm.
While the RW algorithm falls in the middle of the Search and Hard groups in the baseline study, it aligns more closely with the Easy groups in our study. 
This discrepancy can be attributed to the inherent differences between unit and system-level testing. 
Functions can be called directly in unit testing, once a valid instance is created.
%which is conducted as a white-box test.
In contrast, system-level testing requires the use of a proxy method, i.e., data is sent from the user interfaces.
Web APIs differ significantly from object-oriented software in terms of testing. 
In object-oriented programming, operations on an object share the same state by default. 
For example, calls like ``\texttt{x.foo(); x.bar();}'' both operate on the same object ``\texttt{x}''.
However, Web APIs may involve different states when randomizing actions (e.g., the endpoint \texttt{/x/\{id\}} requires a specific ID to reference the same state). 
Consequently, mutation actions occurring in system-level tests are less directed, possibly leading to surfaces with more plateaus.

A similar pattern is observed with the ND parameter. 
In system-level tests, the amount of neutral distance is significantly higher. 
In fact, most of the search occurs in neutral areas, making it challenging to obtain a fit solution.
When examining the neutral volumes of the solutions, we find that results are similar for both types of tests.
However, unit test exhibits a greater neutral volume, allowing the fitness function to guide the search more effectively.

When we analyze the IC metric, we see more information content in unit tests across all groups. 
This indicates that the peak values are higher in unit tests. 
Conversely, the peak number in system-level tests is considerably lower, particularly within the Easy group, where the difference is approximately four times lower, and the information level is significantly lower.
Notably, the information level of the branches where RW is better is almost comparable to that of the other methods.

Regarding the PIC metric, system-level tests show a lower modality level than unit tests.

The most significant difference between the two test types is evident in the DBI metric, which illustrates the distribution of continuous actions.
The DBI metric measures the behaviors that a fitness value performs continuously (Recall Eq.~\ref{eq:density_basin}).
For instance, continuous increases (a sequence of "$11$"), continuous decreases (a sequence of "$\overline{1}\overline{1}$"), or a constant value (a sequence of "$00$") during the search are calculated based on fitness values.
Greater diversity in these sequences results in values approaching $1$, while a lack of diversity (e.g., no change) yields values closer to $0$.
In system-level tests, the diversity of these continuous values is relatively low, with most steps consisting solely of "$00$" sequences, leading to a DBI value near $0$.
In contrast, unit-level tests demonstrate a high level of diversity in these continuous sequences.
Compared to unit testing, fuzzing Web APIs has a much larger search space, and the test suite consists of a greater number of test cases.
Each test typically targets a small segment of code related to the API.
Thus, during fuzzing, each test generated at each step optimizes different branches, while many other branches have a fitness value of 0.
Over the course of the search, a branch may only be affected in a few optimization steps, and since unaffected branches have a fitness value of 0, this results in the DBI for each single branch being close to 0 in our context.

\begin{figure}[t]
	\centering
	\includegraphics[width=1.0\linewidth]{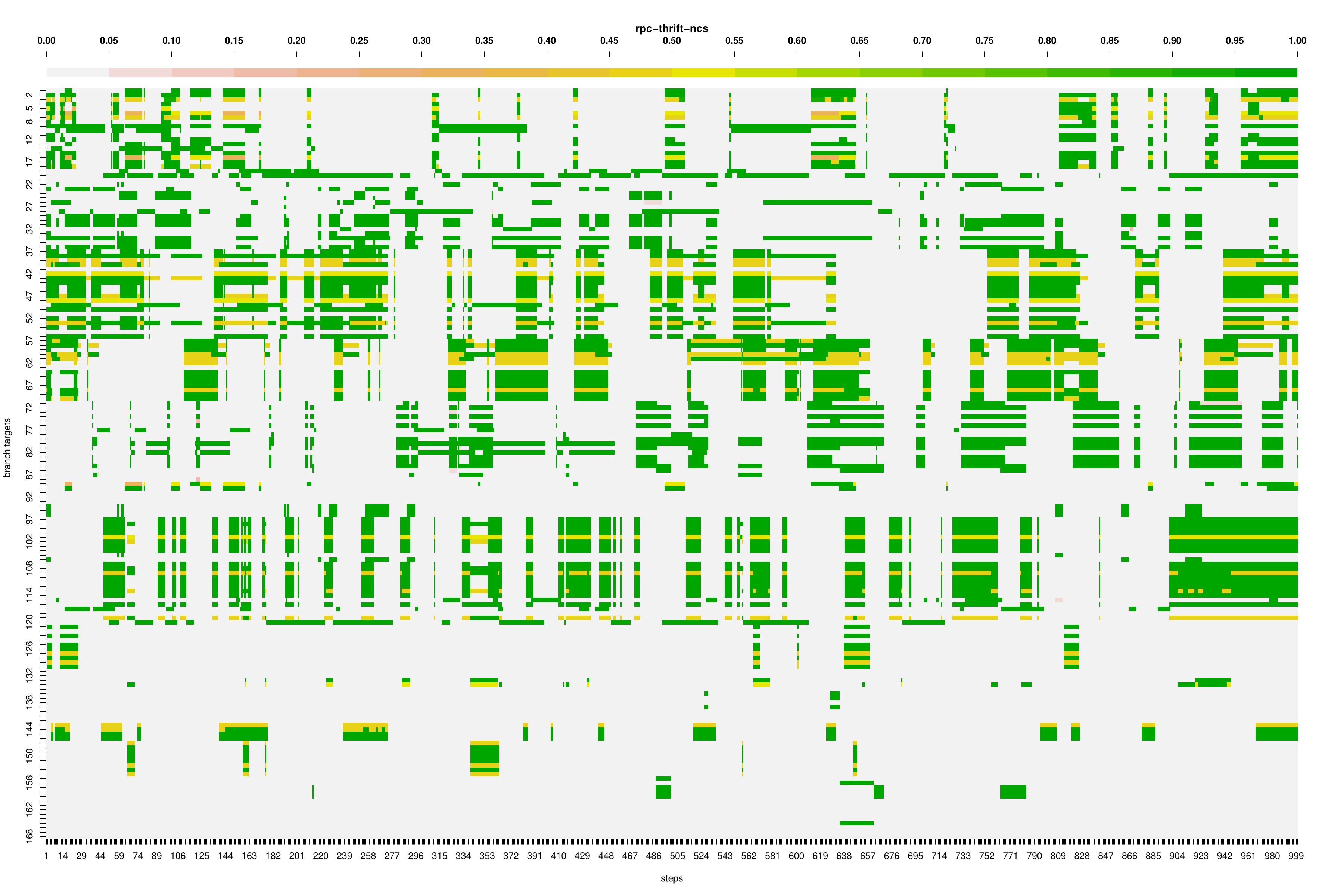}
	\caption{Heatmap representing the fitness values of 168 branches (y-lab is \textit{branches}) tracked over 1000 steps (x-lab is \textit{steps}) during fuzzing with RW on \rpcncs. Note that each row illustrates the fitness values of a single branch at each of 1000 steps.}
		\label{fig:rpc-ncs-dbi}
\end{figure}

To investigate the results of the DBI metric, we analyze the fitness values of 168 branches of \rpcncs (achieved the highest DBI, i.e., $0.089$) over 1000 steps using a heatmap (see Figure~\ref{fig:rpc-ncs-dbi}).
Based on the distribution of fitness values shown in Figure~\ref{fig:rpc-ncs-dbi}, each test generated at a step impacts only a subset of the branches, ranging from 12 ($12/168 = 7.1$\%) to 90 ($90 /168 = 53.6$\%) targets on \rpcncs.
This is because the branches are associated with distinct endpoints of Web APIs and can only be executed if the corresponding endpoint is requested in the test.
As a result, the fitness value for branches unaffected by the test at each step is 0.

Differences between unit and system test generation are excepted.
However, such extreme and sharp difference in DBI values was unexpected, and quite surprising.
The three authors have reviewed all of our code, possibly to see if there was any clear error in our analyses.
For the sake of completeness, we also contacted the original authors of~\cite{albunian2020causes},
asking if the raw data of experiments (missing in the replication package) was still accessible somewhere, of if the data was transformed with some pre-processing technique not discussed in the article and not present in the replication package.
This could have helped in shedding light regarding if there was any error in ours or in their study.
One of such authors replied, unfortunately stating he had no longer access to that data (which is not unreasonable, as few years have passed at the time of writing, and such raw data can take a lot of space).
Still, considering Table~\ref{tab:metrics}, low values for DBI are consistent with the low values we obtained for IC and PIC and the high values for AC, as they all four indicate a lack of ruggedness.
This is what motivated us to wonder whether there was any issue in the handling and interpretation of DBI in~\cite{albunian2020causes}.

\begin{result}
	{\bf RQ4:} System-level fitness landscape characterization includes more plateaus and lower information levels than in unit testing.
\end{result}

%%%%%%%%%%%%%%%%%%%%%%%%%%%%%%%%%%%%%%%%%%%%%%%%%%%%%%%%%%%%%%%%%%%%%%%%%%%%%%%%%%%%%%%%%%%%%%%%%%%%%%%%%%%%%%%
\section{Discussion}
\label{sec:discussion}

First, in this study we compared the RW and MIO algorithms to understand better the differences between both these  algorithms.
As expected, guiding the search with a fitness function yielded positive results in most cases, with the MIO algorithm outperforming the RW algorithm in the majority of the SUTs.
Notably, the MIO algorithm achieved these results while sending significantly fewer HTTP requests than the RW algorithm. 
This suggests that using a fitness function enhances coverage and leads to a more efficient use of resources.

Next, we examined the general characteristics of the fitness landscape. 
Our results were based on a total of~\totalBranches branches, of which~\totalBranchesInFigures were analyzed. 
We excluded~\totalNeverCovered branches that were not covered by either the RW or MIO algorithms during any of their 30 runs.
Most of these branches contain neutral areas of the fitness landscape, with only a tiny portion classified as rugged.

Then, the fitness landscape features of the branches were examined, and their contribution to the success rate was investigated. 
Accordingly, high AC and ND values make generating tests at the system-level difficult. 
The reason for this is that the fitness values have a high level of correlation with each other, which makes it challenging to direct the search. 
In addition, the increase in the ND metric, which is the number of steps between two different fitness values during the search, makes the search difficult, as expected. 
High NV, IC, PIC, and DBI metrics make it easier to generate system-level tests. 
As the NV metric increases, the difference in the number of fitness values in the landscape increases. 
Searching in a landscape with different fitness values will be much easier. 
The increase in the IC metric means that the landscape has higher information. 
While this can sometimes make the search difficult, the increase in information in system-level test generation, where the landscape generally consists of neutral areas, has been quite important for directing the search. 
PIC provides information about the modality of the landscape. 
The effect of single-modal or multimodal problems on the difficulty may vary depending on the type of problem. 
When the effect on system-level test case generation is examined, it can be said that it is easier to cover multimodal branches. 
The search becomes more manageable with the increase of the DBI metric. 
DBI helps us understand the diversity of flat areas. 
It is seen that flat areas negatively affect the search. 
At least the diversification of these flat areas affects the search positively. 
When all these metrics are examined, it is seen that any piece of information that can guide the search in the landscape positively affects the search. 

The next phase of the research focuses on investigating how source code features influence search processes. 
Various types of branches were examined, leading to specific findings. 
It is observed that fitness diversity is significantly lower in groups where the RW algorithm outperforms the MIO algorithm. 
This suggests the presence of branches that cannot be influenced by fitness evaluation and can only be addressed through random mutations. 
One key finding is that most of these branches are of the ``Integer\_Zero'' type. 
This explains the high number of neutral areas present in the search landscape. 
Branches in the ``Integer\_Zero'' category typically arise from the Java compiler's conversion of boolean predicates, and these boolean predicates often lead to plateaus in the fitness landscape~(\cite{Mcm04}).
Additionally, when examining the branches that are never covered, the most common branches are ``Reference\_Null'' branches that follow the branches in the ``Integer\_Zero'' group. These branches cause boolean flags and plateaus in the fitness landscape.
%These branches typically consist of plateaus that emerge after Java bytecode transformation\andrea{what you mean?}.
%Additionally, when examining the branches that are never covered, it becomes evident that there are References to Null comparisons alongside the boolean flags\andrea{isn't the same?}.
This situation often arises from \texttt{null} checks implemented to prevent program crashes or due to the requirements of null-safe programming languages, making it challenging to cover branches associated with these \texttt{null} values.

Finally, the differences between unit and system-level test generation are compared using results obtained from the replicated study.
As in system-level test generation, while the increase in AC and ND values negatively affects the search, the increase in NV, IC, PIC values positively affects the search.
While the values are generally similar, there are some differences.
A possible explanation for these differences is that specific methods can be called directly when generating a unit test.
This allows the fitness function to be used more effectively. 
In system-level test generation, it is realized through proxy methods. 
This makes it challenging to perform the search. 
This may be the main reason for the differences between metric values. 

The most obvious of these differences is the DBI metric.
While it is very close to the value of 1 in unit test generation, it is close to 0 in system-level test generation. 
This is related to the variety of flat areas obtained during the search. 
As in unit testing, the system-level fitness landscape generally consists of flat areas. 
However, the variety of these flat areas is quite limited in system-level testing. 
They mainly consist of ``00'' sequences, meaning no change.
As a result, while the fitness landscape has more plateaus in system-level test generation compared to unit test generation, the level of information is also less.
Alternative, a possible explanation for this huge differences between DBI metric values could be due to possible errors in our analyses or in the analyses carried out in~\cite{albunian2020causes}.

In future studies, it will be possible to 
%determine the region of the landscape where a solution is located and define a fitness function based on that information\andrea{what information?}.
define a fitness function appropriate for a solution's landscape's ruggedness or neutrality characteristics that can help guide the search more effectively.
%Defining new fitness functions that provide insights into generating test cases at the system-level is crucial. 
Additionally, a new solution generation mechanism can be developed. 
For instance, in a landscape characterized by many plateaus, the effectiveness of the crossover operator is limited, making it generally more meaningful to generate new individuals using the mutation operator. 
Besides, solution mechanisms like crossover, which allows two individuals to exchange information, are expected to yield better results in information-rugged landscapes.
Currently, MIO and \evo do not use any kind of crossover operator, due to the complexity of defining a meaningful one for this problem domain.
The results of this study points to the possible need to develop novel mutation and crossover operators, customized for this problem domain, taking into account these fitness landscape characteristics.
Also, in retrospective this can explain existing results.
For example, the introduction of hypermutation in \evo~(\cite{zhang2021adaptive}) led to significantly better results, which can be explained due the neutrality characteristics of this problem's domain fitness landscape.

%%%%%%%%%%%%%%%%%%%%%%%%%%%%%%%%%%%%%%%%%%%%%%%%%%%%%%%%%%%%%%%%%%%%%%%%%%%%%%%%%%%%%%%%%%%%%%%%%%%%%%%%%%%%%%%
\section{Threats To Validity}
\label{sec:threats}

To address issues with internal validity, we carefully tested our \evo implementation with thousands of unit tests and end-to-end tests.
Also, \evo is released as open source, and
anyone who wants to can check its source code.

To address the randomness of the applied algorithms, each experiment was repeated 30 times and appropriate statistical tests were performed following common literature guidelines~(\cite{Hitchhiker14}).

The external validity threats concern whether our results can be generalized to other SUTs and SBST problems.
We conducted our study on $23$ different SUTs prepared for different kinds of APIs, including REST, GraphQL, and Thrift.
This provides a large variety of different kinds of SUTs, which strengthen our results.
However, system-level testing is computationally costly, and a limited number of SUTs can be used to complete the experiments within reasonable time considering an academic context.
In addition, although especially REST APIs are frequently used in industry, it can be challenging to find and set up open-source SUTs suitable for experiments.

%%%%%%%%%%%%%%%%%%%%%%%%%%%%%%%%%%%%%%%%%%%%%%%%%%%%%%%%%%%%%%%%%%%%%%%%%%%%
\section{Conclusions}
\label{sec:conclusions}

An in-depth evaluation of the performance of search algorithms used in system-level test generation requires a thorough understanding of the fitness function landscapes.
This study examined fitness landscapes regarding ruggedness and neutrality, comparing them to the landscape features found in unit test generation. 
This study on system-level testing with \evo is a replication of an existing study on unit-level testing with EvoSuite~(\cite{albunian2020causes}), using $23$ Web APIs for the experiments.

Our results indicate that fitness landscapes are dominantly characterized by neutral regions, such as plateaus, which complicate the search process.
The presence of information content within the landscape can help with the search.
One primary reason for this neutral landscape is the presence of boolean flags.

Our results confirm the existing results reported in the literature for unit testing, albeit with some differences.
Based on these findings, developing new fitness functions to extract more detailed information from the SUTs,  implementing innovative testability transformations and designing novel mutation and crossover operators will be important priorities to enhance the search process.

%%%%%%%%%%%%%%%%%%%%%%%%%%%%%%%%%%%%%%%%%%%%%%%%%%%%%%%%%%%%%%%%%%%%%%%%%%%%
\section*{Data Availability Statement}

\evo is open-source on GitHub,\footnote{https://github.com/WebFuzzing/EvoMaster}
with each new release, such as version 3.4.0, automatically published on Zenodo~(\cite{zenodo340evomaster}).
Likewise, the corpus EMB used for our experiments is open-source,\footnote{https://github.com/WebFuzzing/EMB}
as well as being stored on Zenodo~(\cite{zenodo340emb}).

To enable the replicability of this study, and simplify comparisons,
all the scripts used to setup our experiments and analyze the results are present
in a replication package on Zenodo~(\cite{sahinFitnessReplication2025}).
This includes as well all the raw data resulting from the experiments.

%%%%%%%%%%%%%%%%%%%%%%%%%%%%%%%%%%%%%%%%%%%%%%%%%%%%%%%%%%%%%%%%%%%%%%%%%%%%
\section*{Author Contribution Statement}

The study was led by Omur Sahin, which did the majority of the work.
Andrea Arcuri did the original planning and design of the work.
All authors contributed with the implementation and extension of \evo needed to carry out the experiments.
The first draft of the article was written by Omur Sahin, where the co-authors iterated over the draft to improve it and add any missing parts.
Experiments were run and first analyzed by Omur Sahin.
All authors contributed to the critical analysis of the results.
Man Zhang further reviewed and analyzed discrepancies in the data (e.g., regarding the DBI metric).

%%%%%%%%%%%%%%%%%%%%%%%%%%%%%%%%%%%%%%%%%%%%%%%%%%%%%%%%%%%%%%%%%%%%%%%%%%%%
\section*{Acknowledgments}
Omur Sahin is supported by the TÜBİTAK 2219 International Postdoctoral Research Fellowship Program (Project ID: 1059B192300060). Man Zhang is supported by State Key Laboratory of  Complex \& Critical Software Environment (SKLCCSE, grant No. CCSE-2024ZX-01) and the Fundamental Research Funds for the Central Universities.
Andrea Arcuri is funded by the European Research Council (ERC) under the European Union’s Horizon 2020 research and innovation programme (EAST project, grant agreement No. 864972).

%%%%%%%%%%%%%%%%%%%%%%%%%%%%%%%%%%%%%%%%%%%%%%%%%%%%%%%%%%%%%%%%%%%%%%%%%%%%

\bibliographystyle{ACM-Reference-Format} % this requires ACM-Reference-Format.bst in same folder
%\bibliographystyle{acm}  % this is ancient from 80s, which does not support URL and DOI

%%https://arxiv.org/help/submit_tex#latex
%
% IMPORTANT: For final version arXiv, use generated bbl.
%            In such case, do not use compile.sh to build final pdf, but rather call pdflatex directly, eg
%            pdflatex arxiv

%\input{.bbl}
\bibliography{papers}


\begin{thebibliography}{85}

%%% the Web address.

\ifx \showCODEN    \undefined \def \showCODEN     #1{\unskip}     \fi
\ifx \showDOI      \undefined \def \showDOI       #1{#1}\fi
\ifx \showISBNx    \undefined \def \showISBNx     #1{\unskip}     \fi
\ifx \showISBNxiii \undefined \def \showISBNxiii  #1{\unskip}     \fi
\ifx \showISSN     \undefined \def \showISSN      #1{\unskip}     \fi
\ifx \showLCCN     \undefined \def \showLCCN      #1{\unskip}     \fi
\ifx \shownote     \undefined \def \shownote      #1{#1}          \fi
\ifx \showarticletitle \undefined \def \showarticletitle #1{#1}   \fi
\ifx \showURL      \undefined \def \showURL       {\relax}        \fi
% invisible to TeX
\providecommand\bibfield[2]{#2}
\providecommand\bibinfo[2]{#2}
\providecommand\natexlab[1]{#1}
\providecommand\showeprint[2][]{arXiv:#2}

\bibitem[Albunian et~al\mbox{.}(2020)]%
        {albunian2020causes}
\bibfield{author}{\bibinfo{person}{Nasser Albunian}, \bibinfo{person}{Gordon
  Fraser}, {and} \bibinfo{person}{Dirk Sudholt}.}
  \bibinfo{year}{2020}\natexlab{}.
\newblock \showarticletitle{Causes and effects of fitness landscapes in unit
  test generation}. In \bibinfo{booktitle}{\emph{Proceedings of the 2020
  Genetic and Evolutionary Computation Conference}}.
  \bibinfo{pages}{1204--1212}.
\newblock


\bibitem[Aleti and Moser(2015)]%
        {aleti2015fitness}
\bibfield{author}{\bibinfo{person}{Aldeida Aleti} {and} \bibinfo{person}{Irene
  Moser}.} \bibinfo{year}{2015}\natexlab{}.
\newblock \showarticletitle{Fitness landscape characterisation for constrained
  software architecture optimisation problems}. In
  \bibinfo{booktitle}{\emph{2015 20th International Conference on Engineering
  of Complex Computer Systems (ICECCS)}}. IEEE, \bibinfo{pages}{11--20}.
\newblock


\bibitem[Aleti et~al\mbox{.}(2017)]%
        {aleti2017analysing}
\bibfield{author}{\bibinfo{person}{Aldeida Aleti}, \bibinfo{person}{Irene
  Moser}, {and} \bibinfo{person}{Lars Grunske}.}
  \bibinfo{year}{2017}\natexlab{}.
\newblock \showarticletitle{Analysing the fitness landscape of search-based
  software testing problems}.
\newblock \bibinfo{journal}{\emph{Automated Software Engineering}}
  \bibinfo{volume}{24} (\bibinfo{year}{2017}), \bibinfo{pages}{603--621}.
\newblock


\bibitem[Ali et~al\mbox{.}(2010)]%
        {ABHP09}
\bibfield{author}{\bibinfo{person}{S. Ali}, \bibinfo{person}{L.C. Briand},
  \bibinfo{person}{H. Hemmati}, {and} \bibinfo{person}{R.K. Panesar-Walawege}.}
  \bibinfo{year}{2010}\natexlab{}.
\newblock \showarticletitle{{A systematic review of the application and
  empirical investigation of search-based test-case generation}}.
\newblock \bibinfo{journal}{\emph{IEEE Transactions on Software Engineering
  (TSE)}} \bibinfo{volume}{36}, \bibinfo{number}{6} (\bibinfo{year}{2010}),
  \bibinfo{pages}{742--762}.
\newblock


\bibitem[Alshraideh and Bottaci(2006)]%
        {AlB06}
\bibfield{author}{\bibinfo{person}{M. Alshraideh} {and} \bibinfo{person}{L.
  Bottaci}.} \bibinfo{year}{2006}\natexlab{}.
\newblock \showarticletitle{Search-based software test data generation for
  string data using program-specific search operators}.
\newblock \bibinfo{journal}{\emph{Software Testing, Verification and
  Reliability (STVR)}} \bibinfo{volume}{16}, \bibinfo{number}{3}
  (\bibinfo{year}{2006}), \bibinfo{pages}{175--203}.
\newblock


\bibitem[Arcuri(2017)]%
        {mio2017}
\bibfield{author}{\bibinfo{person}{Andrea Arcuri}.}
  \bibinfo{year}{2017}\natexlab{}.
\newblock \showarticletitle{{Many Independent Objective (MIO) Algorithm for
  Test Suite Generation}}. In \bibinfo{booktitle}{\emph{International Symposium
  on Search Based Software Engineering (SSBSE)}}. \bibinfo{pages}{3--17}.
\newblock


\bibitem[Arcuri(2018)]%
        {arcuri2018test}
\bibfield{author}{\bibinfo{person}{Andrea Arcuri}.}
  \bibinfo{year}{2018}\natexlab{}.
\newblock \showarticletitle{{Test suite generation with the Many Independent
  Objective (MIO) algorithm}}.
\newblock \bibinfo{journal}{\emph{Information and Software Technology}}
  \bibinfo{volume}{104} (\bibinfo{year}{2018}), \bibinfo{pages}{195--206}.
\newblock


\bibitem[Arcuri(2019)]%
        {arcuri2019restful}
\bibfield{author}{\bibinfo{person}{Andrea Arcuri}.}
  \bibinfo{year}{2019}\natexlab{}.
\newblock \showarticletitle{RESTful API Automated Test Case Generation with
  EvoMaster}.
\newblock \bibinfo{journal}{\emph{ACM Transactions on Software Engineering and
  Methodology (TOSEM)}} \bibinfo{volume}{28}, \bibinfo{number}{1}
  (\bibinfo{year}{2019}), \bibinfo{pages}{3}.
\newblock


\bibitem[Arcuri(2020)]%
        {arcuri2020blackbox}
\bibfield{author}{\bibinfo{person}{Andrea Arcuri}.}
  \bibinfo{year}{2020}\natexlab{}.
\newblock \showarticletitle{Automated Black-and White-Box Testing of RESTful
  APIs With EvoMaster}.
\newblock \bibinfo{journal}{\emph{IEEE Software}} \bibinfo{volume}{38},
  \bibinfo{number}{3} (\bibinfo{year}{2020}), \bibinfo{pages}{72--78}.
\newblock


\bibitem[Arcuri and Briand(2014)]%
        {Hitchhiker14}
\bibfield{author}{\bibinfo{person}{A. Arcuri} {and} \bibinfo{person}{L.
  Briand}.} \bibinfo{year}{2014}\natexlab{}.
\newblock \showarticletitle{{A Hitchhiker's Guide to Statistical Tests for
  Assessing Randomized Algorithms in Software Engineering}}.
\newblock \bibinfo{journal}{\emph{Software Testing, Verification and
  Reliability (STVR)}} \bibinfo{volume}{24}, \bibinfo{number}{3}
  (\bibinfo{year}{2014}), \bibinfo{pages}{219--250}.
\newblock


\bibitem[Arcuri and Fraser(2013)]%
        {arcuri2013parameter}
\bibfield{author}{\bibinfo{person}{Andrea Arcuri} {and} \bibinfo{person}{Gordon
  Fraser}.} \bibinfo{year}{2013}\natexlab{}.
\newblock \showarticletitle{Parameter tuning or default values? An empirical
  investigation in search-based software engineering}.
\newblock \bibinfo{journal}{\emph{Empirical Software Engineering (EMSE)}}
  \bibinfo{volume}{18}, \bibinfo{number}{3} (\bibinfo{year}{2013}),
  \bibinfo{pages}{594--623}.
\newblock


\bibitem[Arcuri and Galeotti(2020a)]%
        {arcuri2020sql}
\bibfield{author}{\bibinfo{person}{Andrea Arcuri} {and} \bibinfo{person}{Juan~P
  Galeotti}.} \bibinfo{year}{2020}\natexlab{a}.
\newblock \showarticletitle{Handling SQL databases in automated system test
  generation}.
\newblock \bibinfo{journal}{\emph{ACM Transactions on Software Engineering and
  Methodology (TOSEM)}} \bibinfo{volume}{29}, \bibinfo{number}{4}
  (\bibinfo{year}{2020}), \bibinfo{pages}{1--31}.
\newblock


\bibitem[Arcuri and Galeotti(2020b)]%
        {arcuri2020testability}
\bibfield{author}{\bibinfo{person}{Andrea Arcuri} {and} \bibinfo{person}{Juan~P
  Galeotti}.} \bibinfo{year}{2020}\natexlab{b}.
\newblock \showarticletitle{Testability transformations for existing APIs}. In
  \bibinfo{booktitle}{\emph{2020 IEEE 13th International Conference on Software
  Testing, Validation and Verification (ICST)}}. IEEE,
  \bibinfo{pages}{153--163}.
\newblock


\bibitem[Arcuri and Galeotti(2021a)]%
        {arcuri2021tt}
\bibfield{author}{\bibinfo{person}{Andrea Arcuri} {and} \bibinfo{person}{Juan~P
  Galeotti}.} \bibinfo{year}{2021}\natexlab{a}.
\newblock \showarticletitle{Enhancing Search-based Testing with Testability
  Transformations for Existing APIs}.
\newblock \bibinfo{journal}{\emph{ACM Transactions on Software Engineering and
  Methodology (TOSEM)}} \bibinfo{volume}{31}, \bibinfo{number}{1}
  (\bibinfo{year}{2021}), \bibinfo{pages}{1--34}.
\newblock


\bibitem[Arcuri and Galeotti(2021b)]%
        {arcuri2021enhancing}
\bibfield{author}{\bibinfo{person}{Andrea Arcuri} {and} \bibinfo{person}{Juan~P
  Galeotti}.} \bibinfo{year}{2021}\natexlab{b}.
\newblock \showarticletitle{{Enhancing Search-based Testing with Testability
  Transformations for Existing APIs}}.
\newblock \bibinfo{journal}{\emph{ACM Transactions on Software Engineering and
  Methodology (TOSEM)}} \bibinfo{volume}{31}, \bibinfo{number}{1}
  (\bibinfo{year}{2021}), \bibinfo{pages}{1--34}.
\newblock


\bibitem[Arcuri et~al\mbox{.}(2021)]%
        {arcuri2021evomaster}
\bibfield{author}{\bibinfo{person}{Andrea Arcuri}, \bibinfo{person}{Juan~Pablo
  Galeotti}, \bibinfo{person}{Bogdan Marculescu}, {and} \bibinfo{person}{Man
  Zhang}.} \bibinfo{year}{2021}\natexlab{}.
\newblock \showarticletitle{EvoMaster: A Search-Based System Test Generation
  Tool}.
\newblock \bibinfo{journal}{\emph{Journal of Open Source Software}}
  \bibinfo{volume}{6}, \bibinfo{number}{57} (\bibinfo{year}{2021}),
  \bibinfo{pages}{2153}.
\newblock


\bibitem[Arcuri et~al\mbox{.}(2012)]%
        {AIB11}
\bibfield{author}{\bibinfo{person}{A. Arcuri}, \bibinfo{person}{M.~Z. Iqbal},
  {and} \bibinfo{person}{L. Briand}.} \bibinfo{year}{2012}\natexlab{}.
\newblock \showarticletitle{Random Testing: Theoretical Results and Practical
  Implications}.
\newblock \bibinfo{journal}{\emph{IEEE Transactions on Software Engineering
  (TSE)}} \bibinfo{volume}{38}, \bibinfo{number}{2} (\bibinfo{year}{2012}),
  \bibinfo{pages}{258--277}.
\newblock


\bibitem[Arcuri et~al\mbox{.}(2025a)]%
        {icst2025vw}
\bibfield{author}{\bibinfo{person}{A. Arcuri}, \bibinfo{person}{A. Poth}, {and}
  \bibinfo{person}{O. Rrjolli}.} \bibinfo{year}{2025}\natexlab{a}.
\newblock \showarticletitle{Introducing Black-Box Fuzz Testing for REST APIs in
  Industry: Challenges and Solutions}. In \bibinfo{booktitle}{\emph{IEEE
  International Conference on Software Testing, Verification and Validation
  (ICST)}}.
\newblock


\bibitem[Arcuri et~al\mbox{.}(2023a)]%
        {arcuri2023advanced}
\bibfield{author}{\bibinfo{person}{Andrea Arcuri}, \bibinfo{person}{Man Zhang},
  {and} \bibinfo{person}{Juan~Pablo Galeotti}.}
  \bibinfo{year}{2023}\natexlab{a}.
\newblock \showarticletitle{Advanced White-Box Heuristics for Search-Based
  Fuzzing of REST APIs}.
\newblock \bibinfo{journal}{\emph{arXiv preprint arXiv:2309.08360}}
  (\bibinfo{year}{2023}).
\newblock


\bibitem[Arcuri et~al\mbox{.}(2025b)]%
        {zenodo340emb}
\bibfield{author}{\bibinfo{person}{Andrea Arcuri}, \bibinfo{person}{Man Zhang},
  \bibinfo{person}{Amid Golmohammadi}, \bibinfo{person}{Asma Belhadi},
  \bibinfo{person}{Onur Duman}, \bibinfo{person}{Susruthan Seran},
  \bibinfo{person}{Juan~Pablo Galeotti}, {and} \bibinfo{person}{Hernan
  Ghianni}.} \bibinfo{year}{2025}\natexlab{b}.
\newblock \bibinfo{booktitle}{\emph{WebFuzzing/EMB: v3.4.0}}.
\newblock
\urldef\tempurl%
\url{https://doi.org/10.5281/zenodo.14597431}
\showDOI{\tempurl}


\bibitem[Arcuri et~al\mbox{.}(2023b)]%
        {icst2023emb}
\bibfield{author}{\bibinfo{person}{Andrea Arcuri}, \bibinfo{person}{Man Zhang},
  \bibinfo{person}{Amid Golmohammadi}, \bibinfo{person}{Asma Belhadi},
  \bibinfo{person}{Juan~P Galeotti}, \bibinfo{person}{Bogdan Marculescu}, {and}
  \bibinfo{person}{Susruthan Seran}.} \bibinfo{year}{2023}\natexlab{b}.
\newblock \showarticletitle{Emb: A curated corpus of web/enterprise
  applications and library support for software testing research}. In
  \bibinfo{booktitle}{\emph{2023 IEEE Conference on Software Testing,
  Verification and Validation (ICST)}}. IEEE, \bibinfo{pages}{433--442}.
\newblock


\bibitem[Arcuri et~al\mbox{.}(2025c)]%
        {zenodo340evomaster}
\bibfield{author}{\bibinfo{person}{Andrea Arcuri}, \bibinfo{person}{Man Zhang},
  \bibinfo{person}{Susruthan Seran}, \bibinfo{person}{Asma Belhadi},
  \bibinfo{person}{Juan~Pablo Galeotti}, \bibinfo{person}{Bogdan},
  \bibinfo{person}{Amid Golmohammadi}, \bibinfo{person}{Onur Duman},
  \bibinfo{person}{Agustina Aldasoro}, \bibinfo{person}{Philip},
  \bibinfo{person}{Alberto~Martín López}, \bibinfo{person}{Hernan Ghianni},
  \bibinfo{person}{Ömür Şahin}, \bibinfo{person}{Annibale Panichella},
  \bibinfo{person}{Kyle Niemeyer}, {and} \bibinfo{person}{Marcello Maugeri}.}
  \bibinfo{year}{2025}\natexlab{c}.
\newblock \bibinfo{booktitle}{\emph{WebFuzzing/EvoMaster: v3.4.0}}.
\newblock
\urldef\tempurl%
\url{https://doi.org/10.5281/zenodo.14597412}
\showDOI{\tempurl}


\bibitem[Arcuri et~al\mbox{.}(2025d)]%
        {arcuri2025tool}
\bibfield{author}{\bibinfo{person}{Andrea Arcuri}, \bibinfo{person}{Man Zhang},
  \bibinfo{person}{Susruthan Seran}, \bibinfo{person}{Juan~Pablo Galeotti},
  \bibinfo{person}{Amid Golmohammadi}, \bibinfo{person}{Onur Duman},
  \bibinfo{person}{Agustina Aldasoro}, {and} \bibinfo{person}{Hernan Ghianni}.}
  \bibinfo{year}{2025}\natexlab{d}.
\newblock \showarticletitle{Tool report: EvoMaster—black and white box
  search-based fuzzing for REST, GraphQL and RPC APIs}.
\newblock \bibinfo{journal}{\emph{Automated Software Engineering}}
  \bibinfo{volume}{32}, \bibinfo{number}{1} (\bibinfo{year}{2025}),
  \bibinfo{pages}{1--11}.
\newblock


\bibitem[Atlidakis et~al\mbox{.}(2019)]%
        {restlerICSE2019}
\bibfield{author}{\bibinfo{person}{Vaggelis Atlidakis},
  \bibinfo{person}{Patrice Godefroid}, {and} \bibinfo{person}{Marina
  Polishchuk}.} \bibinfo{year}{2019}\natexlab{}.
\newblock \showarticletitle{RESTler: Stateful {REST} {API} Fuzzing}. In
  \bibinfo{booktitle}{\emph{ACM/IEEE International Conference on Software
  Engineering (ICSE)}}. \bibinfo{pages}{748–758}.
\newblock


\bibitem[Barnett et~al\mbox{.}(1998)]%
        {barnett1998ruggedness}
\bibfield{author}{\bibinfo{person}{Lionel Barnett} {et~al\mbox{.}}}
  \bibinfo{year}{1998}\natexlab{}.
\newblock \showarticletitle{Ruggedness and neutrality-the NKp family of fitness
  landscapes}. In \bibinfo{booktitle}{\emph{Artificial Life VI: Proceedings of
  the sixth international conference on Artificial life}}.
  \bibinfo{pages}{18--27}.
\newblock


\bibitem[Belhadi et~al\mbox{.}(2023)]%
        {belhadi2023random}
\bibfield{author}{\bibinfo{person}{Asma Belhadi}, \bibinfo{person}{Man Zhang},
  {and} \bibinfo{person}{Andrea Arcuri}.} \bibinfo{year}{2023}\natexlab{}.
\newblock \showarticletitle{Random Testing and Evolutionary Testing for Fuzzing
  GraphQL APIs}.
\newblock \bibinfo{journal}{\emph{ACM Transactions on the Web}}
  (\bibinfo{year}{2023}).
\newblock


\bibitem[Bertolino(2007)]%
        {bertolino2007software}
\bibfield{author}{\bibinfo{person}{A. Bertolino}.}
  \bibinfo{year}{2007}\natexlab{}.
\newblock \showarticletitle{Software testing research: Achievements,
  challenges, dreams}. In \bibinfo{booktitle}{\emph{Future of Software
  Engineering, 2007. FOSE'07}}. IEEE, \bibinfo{pages}{85--103}.
\newblock


\bibitem[Bozkurt et~al\mbox{.}(2013)]%
        {bozkurt2013testing}
\bibfield{author}{\bibinfo{person}{Mustafa Bozkurt}, \bibinfo{person}{Mark
  Harman}, {and} \bibinfo{person}{Youssef Hassoun}.}
  \bibinfo{year}{2013}\natexlab{}.
\newblock \showarticletitle{Testing and verification in service-oriented
  architecture: a survey}.
\newblock \bibinfo{journal}{\emph{Software Testing, Verification and
  Reliability (STVR)}} \bibinfo{volume}{23}, \bibinfo{number}{4}
  (\bibinfo{year}{2013}), \bibinfo{pages}{261--313}.
\newblock


\bibitem[Campos et~al\mbox{.}(2018)]%
        {campos2018empirical}
\bibfield{author}{\bibinfo{person}{Jos{\'e} Campos}, \bibinfo{person}{Yan Ge},
  \bibinfo{person}{Nasser Albunian}, \bibinfo{person}{Gordon Fraser},
  \bibinfo{person}{Marcelo Eler}, {and} \bibinfo{person}{Andrea Arcuri}.}
  \bibinfo{year}{2018}\natexlab{}.
\newblock \showarticletitle{An empirical evaluation of evolutionary algorithms
  for unit test suite generation}.
\newblock \bibinfo{journal}{\emph{Information and Software Technology (IST)}}
  \bibinfo{volume}{104} (\bibinfo{year}{2018}), \bibinfo{pages}{207--235}.
\newblock


\bibitem[Canfora and Di~Penta(2009)]%
        {canfora2009service}
\bibfield{author}{\bibinfo{person}{Gerardo Canfora} {and}
  \bibinfo{person}{Massimiliano Di~Penta}.} \bibinfo{year}{2009}\natexlab{}.
\newblock \showarticletitle{{Service-oriented architectures testing: A
  survey}}.
\newblock In \bibinfo{booktitle}{\emph{Software Engineering}}.
  \bibinfo{publisher}{Springer}, \bibinfo{pages}{78--105}.
\newblock


\bibitem[Chicano et~al\mbox{.}(2011)]%
        {chicano2011elementary}
\bibfield{author}{\bibinfo{person}{Francisco Chicano}, \bibinfo{person}{Javier
  Ferrer}, {and} \bibinfo{person}{Enrique Alba}.}
  \bibinfo{year}{2011}\natexlab{}.
\newblock \showarticletitle{Elementary landscape decomposition of the test
  suite minimization problem}. In \bibinfo{booktitle}{\emph{Search Based
  Software Engineering: Third International Symposium, SSBSE 2011, Szeged,
  Hungary, September 10-12, 2011. Proceedings 3}}. Springer,
  \bibinfo{pages}{48--63}.
\newblock


\bibitem[Corradini et~al\mbox{.}(2024)]%
        {corradini2024deeprest}
\bibfield{author}{\bibinfo{person}{Davide Corradini}, \bibinfo{person}{Zeno
  Montolli}, \bibinfo{person}{Michele Pasqua}, {and} \bibinfo{person}{Mariano
  Ceccato}.} \bibinfo{year}{2024}\natexlab{}.
\newblock \showarticletitle{DeepREST: Automated Test Case Generation for REST
  APIs Exploiting Deep Reinforcement Learning}.
\newblock \bibinfo{journal}{\emph{arXiv preprint arXiv:2408.08594}}
  (\bibinfo{year}{2024}).
\newblock


\bibitem[Curbera et~al\mbox{.}(2002)]%
        {curbera2002unraveling}
\bibfield{author}{\bibinfo{person}{Francisco Curbera}, \bibinfo{person}{Matthew
  Duftler}, \bibinfo{person}{Rania Khalaf}, \bibinfo{person}{William Nagy},
  \bibinfo{person}{Nirmal Mukhi}, {and} \bibinfo{person}{Sanjiva Weerawarana}.}
  \bibinfo{year}{2002}\natexlab{}.
\newblock \showarticletitle{Unraveling the Web services web: an introduction to
  SOAP, WSDL, and UDDI}.
\newblock \bibinfo{journal}{\emph{IEEE Internet computing}}
  \bibinfo{volume}{6}, \bibinfo{number}{2} (\bibinfo{year}{2002}),
  \bibinfo{pages}{86--93}.
\newblock


\bibitem[Droste et~al\mbox{.}(2002)]%
        {DJW02}
\bibfield{author}{\bibinfo{person}{S. Droste}, \bibinfo{person}{T. Jansen},
  {and} \bibinfo{person}{I. Wegener}.} \bibinfo{year}{2002}\natexlab{}.
\newblock \showarticletitle{On the analysis of the (1+1) evolutionary
  algorithm}.
\newblock \bibinfo{journal}{\emph{Theoretical Computer Science}}
  \bibinfo{volume}{276} (\bibinfo{year}{2002}), \bibinfo{pages}{51--81}.
\newblock


\bibitem[Duran and Ntafos(1984)]%
        {DuN84}
\bibfield{author}{\bibinfo{person}{J.~W. Duran} {and} \bibinfo{person}{S.~C.
  Ntafos}.} \bibinfo{year}{1984}\natexlab{}.
\newblock \showarticletitle{An Evaluation of Random Testing}.
\newblock \bibinfo{journal}{\emph{IEEE Transactions on Software Engineering
  (TSE)}} \bibinfo{volume}{10}, \bibinfo{number}{4} (\bibinfo{year}{1984}),
  \bibinfo{pages}{438--444}.
\newblock


\bibitem[Fielding(2000)]%
        {fielding2000architectural}
\bibfield{author}{\bibinfo{person}{Roy~Thomas Fielding}.}
  \bibinfo{year}{2000}\natexlab{}.
\newblock \emph{\bibinfo{title}{Architectural styles and the design of
  network-based software architectures}}.
\newblock \bibinfo{thesistype}{Ph.\,D. Dissertation}.
  \bibinfo{school}{University of California, Irvine}.
\newblock


\bibitem[Fraser and Arcuri(2011)]%
        {fraser2011evosuite}
\bibfield{author}{\bibinfo{person}{Gordon Fraser} {and} \bibinfo{person}{Andrea
  Arcuri}.} \bibinfo{year}{2011}\natexlab{}.
\newblock \showarticletitle{Evo{S}uite: automatic generation for
  object-oriented software}. In \bibinfo{booktitle}{\emph{ACM Symposium on the
  Foundations of Software Engineering (FSE)}}. \bibinfo{pages}{416--419}.
\newblock


\bibitem[Fraser and Arcuri(2013a)]%
        {evosuiteAtSbst2013}
\bibfield{author}{\bibinfo{person}{Gordon Fraser} {and} \bibinfo{person}{Andrea
  Arcuri}.} \bibinfo{year}{2013}\natexlab{a}.
\newblock \showarticletitle{EvoSuite at the {SBST} 2013 Tool Competition}. In
  \bibinfo{booktitle}{\emph{International Workshop on Search-Based Software
  Testing (SBST)}}. \bibinfo{pages}{406--409}.
\newblock


\bibitem[Fraser and Arcuri(2013b)]%
        {GoA_TSE12}
\bibfield{author}{\bibinfo{person}{Gordon Fraser} {and} \bibinfo{person}{Andrea
  Arcuri}.} \bibinfo{year}{2013}\natexlab{b}.
\newblock \showarticletitle{Whole Test Suite Generation}.
\newblock \bibinfo{journal}{\emph{IEEE Transactions on Software Engineering}}
  \bibinfo{volume}{39}, \bibinfo{number}{2} (\bibinfo{year}{2013}),
  \bibinfo{pages}{276--291}.
\newblock
\showISSN{0098-5589}


\bibitem[Gallagher and Narasimhan(1997)]%
        {gallagher1997adtest}
\bibfield{author}{\bibinfo{person}{Matthew~J. Gallagher} {and}
  \bibinfo{person}{V~Lakshmi Narasimhan}.} \bibinfo{year}{1997}\natexlab{}.
\newblock \showarticletitle{Adtest: A test data generation suite for ada
  software systems}.
\newblock \bibinfo{journal}{\emph{IEEE Transactions on Software Engineering
  (TSE)}} \bibinfo{volume}{23}, \bibinfo{number}{8} (\bibinfo{year}{1997}),
  \bibinfo{pages}{473--484}.
\newblock


\bibitem[Golmohammadi et~al\mbox{.}(2023a)]%
        {golmohammadi2023net}
\bibfield{author}{\bibinfo{person}{Amid Golmohammadi}, \bibinfo{person}{Man
  Zhang}, {and} \bibinfo{person}{Andrea Arcuri}.}
  \bibinfo{year}{2023}\natexlab{a}.
\newblock \showarticletitle{.{NET}/{C}\# instrumentation for search-based
  software testing}.
\newblock \bibinfo{journal}{\emph{Software Quality Journal}}
  (\bibinfo{year}{2023}), \bibinfo{pages}{1--27}.
\newblock


\bibitem[Golmohammadi et~al\mbox{.}(2023b)]%
        {golmohammadi2023impact}
\bibfield{author}{\bibinfo{person}{Amid Golmohammadi}, \bibinfo{person}{Man
  Zhang}, {and} \bibinfo{person}{Andrea Arcuri}.}
  \bibinfo{year}{2023}\natexlab{b}.
\newblock \showarticletitle{On the Impact of Tool Evolution and Case Study Size
  on SBSE Experiments: A Replicated Study with EvoMaster}. In
  \bibinfo{booktitle}{\emph{International Symposium on Search Based Software
  Engineering}}. Springer, \bibinfo{pages}{108--122}.
\newblock


\bibitem[Golmohammadi et~al\mbox{.}(2023c)]%
        {golmohammadi2023testing}
\bibfield{author}{\bibinfo{person}{Amid Golmohammadi}, \bibinfo{person}{Man
  Zhang}, {and} \bibinfo{person}{Andrea Arcuri}.}
  \bibinfo{year}{2023}\natexlab{c}.
\newblock \showarticletitle{Testing RESTful APIs: A Survey}.
\newblock \bibinfo{journal}{\emph{ACM Transactions on Software Engineering and
  Methodology}} (\bibinfo{date}{aug} \bibinfo{year}{2023}).
\newblock
\showISSN{1049-331X}
\urldef\tempurl%
\url{https://doi.org/10.1145/3617175}
\showDOI{\tempurl}


\bibitem[Harman et~al\mbox{.}(2002)]%
        {HHH02}
\bibfield{author}{\bibinfo{person}{M. Harman}, \bibinfo{person}{L. Hu},
  \bibinfo{person}{R. Hierons}, \bibinfo{person}{A. Baresel}, {and}
  \bibinfo{person}{H. Sthamer}.} \bibinfo{year}{2002}\natexlab{}.
\newblock \showarticletitle{Improving evolutionary testing by flag removal}. In
  \bibinfo{booktitle}{\emph{Genetic and Evolutionary Computation Conference
  (GECCO)}}. \bibinfo{pages}{1351--1358}.
\newblock


\bibitem[Harman et~al\mbox{.}(2004)]%
        {HHH04}
\bibfield{author}{\bibinfo{person}{M. Harman}, \bibinfo{person}{L. Hu},
  \bibinfo{person}{R. Hierons}, \bibinfo{person}{J. Wegener},
  \bibinfo{person}{H. Sthamer}, \bibinfo{person}{A. Baresel}, {and}
  \bibinfo{person}{M. Roper}.} \bibinfo{year}{2004}\natexlab{}.
\newblock \showarticletitle{Testability Transformation}.
\newblock \bibinfo{journal}{\emph{IEEE Transactions on Software Engineering}}
  \bibinfo{volume}{30}, \bibinfo{number}{1} (\bibinfo{year}{2004}),
  \bibinfo{pages}{3--16}.
\newblock


\bibitem[Harman and Jones(2001)]%
        {HaJ01}
\bibfield{author}{\bibinfo{person}{M. Harman} {and} \bibinfo{person}{B.~F.
  Jones}.} \bibinfo{year}{2001}\natexlab{}.
\newblock \showarticletitle{Search-based software engineering}.
\newblock \bibinfo{journal}{\emph{Journal of Information \& Software
  Technology}} \bibinfo{volume}{43}, \bibinfo{number}{14}
  (\bibinfo{year}{2001}), \bibinfo{pages}{833--839}.
\newblock


\bibitem[Harman et~al\mbox{.}(2012)]%
        {harman2012search}
\bibfield{author}{\bibinfo{person}{Mark Harman}, \bibinfo{person}{S~Afshin
  Mansouri}, {and} \bibinfo{person}{Yuanyuan Zhang}.}
  \bibinfo{year}{2012}\natexlab{}.
\newblock \showarticletitle{Search-based software engineering: Trends,
  techniques and applications}.
\newblock \bibinfo{journal}{\emph{ACM Computing Surveys (CSUR)}}
  \bibinfo{volume}{45}, \bibinfo{number}{1} (\bibinfo{year}{2012}),
  \bibinfo{pages}{11}.
\newblock


\bibitem[Hatfield-Dodds and Dygalo(2022)]%
        {hatfield2022deriving}
\bibfield{author}{\bibinfo{person}{Zac Hatfield-Dodds} {and}
  \bibinfo{person}{Dmitry Dygalo}.} \bibinfo{year}{2022}\natexlab{}.
\newblock \showarticletitle{Deriving Semantics-Aware Fuzzers from Web API
  Schemas}. In \bibinfo{booktitle}{\emph{2022 IEEE/ACM 44th International
  Conference on Software Engineering: Companion Proceedings (ICSE-Companion)}}.
  IEEE, \bibinfo{pages}{345--346}.
\newblock


\bibitem[Holland(1992)]%
        {holland1992genetic}
\bibfield{author}{\bibinfo{person}{John~H Holland}.}
  \bibinfo{year}{1992}\natexlab{}.
\newblock \showarticletitle{Genetic algorithms}.
\newblock \bibinfo{journal}{\emph{Scientific american}} \bibinfo{volume}{267},
  \bibinfo{number}{1} (\bibinfo{year}{1992}), \bibinfo{pages}{66--73}.
\newblock


\bibitem[Kim et~al\mbox{.}(2023)]%
        {kim2023adaptive}
\bibfield{author}{\bibinfo{person}{Myeongsoo Kim}, \bibinfo{person}{Saurabh
  Sinha}, {and} \bibinfo{person}{Alessandro Orso}.}
  \bibinfo{year}{2023}\natexlab{}.
\newblock \showarticletitle{Adaptive rest api testing with reinforcement
  learning}. In \bibinfo{booktitle}{\emph{2023 38th IEEE/ACM International
  Conference on Automated Software Engineering (ASE)}}. IEEE,
  \bibinfo{pages}{446--458}.
\newblock


\bibitem[Korel(1990)]%
        {Kor90}
\bibfield{author}{\bibinfo{person}{Bogdan Korel}.}
  \bibinfo{year}{1990}\natexlab{}.
\newblock \showarticletitle{Automated software test data generation}.
\newblock \bibinfo{journal}{\emph{IEEE Transactions on software engineering}}
  \bibinfo{volume}{16}, \bibinfo{number}{8} (\bibinfo{year}{1990}),
  \bibinfo{pages}{870--879}.
\newblock


\bibitem[Kotelyanskii and Kapfhammer(2014)]%
        {kotelyanskii2014parameter}
\bibfield{author}{\bibinfo{person}{Anton Kotelyanskii} {and}
  \bibinfo{person}{Gregory~M Kapfhammer}.} \bibinfo{year}{2014}\natexlab{}.
\newblock \showarticletitle{Parameter tuning for search-based test-data
  generation revisited: Support for previous results}. In
  \bibinfo{booktitle}{\emph{2014 14th International Conference on Quality
  Software}}. IEEE, \bibinfo{pages}{79--84}.
\newblock


\bibitem[Laranjeiro et~al\mbox{.}(2021)]%
        {laranjeiro2021black}
\bibfield{author}{\bibinfo{person}{Nuno Laranjeiro}, \bibinfo{person}{Jo{\~a}o
  Agnelo}, {and} \bibinfo{person}{Jorge Bernardino}.}
  \bibinfo{year}{2021}\natexlab{}.
\newblock \showarticletitle{A black box tool for robustness testing of REST
  services}.
\newblock \bibinfo{journal}{\emph{IEEE Access}}  \bibinfo{volume}{9}
  (\bibinfo{year}{2021}), \bibinfo{pages}{24738--24754}.
\newblock


\bibitem[Lefticaru and Ipate(2008)]%
        {lefticaru2008comparative}
\bibfield{author}{\bibinfo{person}{Raluca Lefticaru} {and}
  \bibinfo{person}{Florentin Ipate}.} \bibinfo{year}{2008}\natexlab{}.
\newblock \showarticletitle{A comparative landscape analysis of fitness
  functions for search-based testing}. In \bibinfo{booktitle}{\emph{2008 10th
  international symposium on symbolic and numeric algorithms for scientific
  computing}}. IEEE, \bibinfo{pages}{201--208}.
\newblock


\bibitem[Liu et~al\mbox{.}(2022)]%
        {liu2022icse}
\bibfield{author}{\bibinfo{person}{Yi Liu}, \bibinfo{person}{Yuekang Li},
  \bibinfo{person}{Gelei Deng}, \bibinfo{person}{Yang Liu},
  \bibinfo{person}{Ruiyuan Wan}, \bibinfo{person}{Runchao Wu},
  \bibinfo{person}{Dandan Ji}, \bibinfo{person}{Shiheng Xu}, {and}
  \bibinfo{person}{Minli Bao}.} \bibinfo{year}{2022}\natexlab{}.
\newblock \showarticletitle{Morest: Model-based RESTful API Testing with
  Execution Feedback}. In \bibinfo{booktitle}{\emph{ACM/IEEE International
  Conference on Software Engineering (ICSE)}}.
\newblock


\bibitem[Lu et~al\mbox{.}(2010)]%
        {LBY10}
\bibfield{author}{\bibinfo{person}{G. Lu}, \bibinfo{person}{R. Bahsoon}, {and}
  \bibinfo{person}{X. Yao}.} \bibinfo{year}{2010}\natexlab{}.
\newblock \showarticletitle{{Applying Elementary Landscape Analysis to
  Search-Based Software Engineering}}. In
  \bibinfo{booktitle}{\emph{International Symposium on Search Based Software
  Engineering (SSBSE)}}. \bibinfo{pages}{3--8}.
\newblock


\bibitem[Lukasczyk and Fraser(2022)]%
        {lukasczyk2022pynguin}
\bibfield{author}{\bibinfo{person}{Stephan Lukasczyk} {and}
  \bibinfo{person}{Gordon Fraser}.} \bibinfo{year}{2022}\natexlab{}.
\newblock \showarticletitle{Pynguin: Automated unit test generation for
  python}. In \bibinfo{booktitle}{\emph{Proceedings of the ACM/IEEE 44th
  International Conference on Software Engineering: Companion Proceedings}}.
  \bibinfo{pages}{168--172}.
\newblock


\bibitem[Malan and Engelbrecht(2013)]%
        {malan2013survey}
\bibfield{author}{\bibinfo{person}{Katherine~M Malan} {and}
  \bibinfo{person}{Andries~P Engelbrecht}.} \bibinfo{year}{2013}\natexlab{}.
\newblock \showarticletitle{A survey of techniques for characterising fitness
  landscapes and some possible ways forward}.
\newblock \bibinfo{journal}{\emph{Information Sciences}}  \bibinfo{volume}{241}
  (\bibinfo{year}{2013}), \bibinfo{pages}{148--163}.
\newblock


\bibitem[Mao et~al\mbox{.}(2016)]%
        {mao2016sapienz}
\bibfield{author}{\bibinfo{person}{Ke Mao}, \bibinfo{person}{Mark Harman},
  {and} \bibinfo{person}{Yue Jia}.} \bibinfo{year}{2016}\natexlab{}.
\newblock \showarticletitle{{Sapienz: Multi-objective automated testing for
  android applications}}. In \bibinfo{booktitle}{\emph{ACM Int. Symposium on
  Software Testing and Analysis (ISSTA)}}. ACM, \bibinfo{pages}{94--105}.
\newblock


\bibitem[Martin-Lopez et~al\mbox{.}(2021)]%
        {martinLopez2021Restest}
\bibfield{author}{\bibinfo{person}{Alberto Martin-Lopez},
  \bibinfo{person}{Sergio Segura}, {and} \bibinfo{person}{Antonio
  Ruiz-Cort\'{e}s}.} \bibinfo{year}{2021}\natexlab{}.
\newblock \showarticletitle{{RESTest: Automated Black-Box Testing of RESTful
  Web APIs}}. In \bibinfo{booktitle}{\emph{ACM Int. Symposium on Software
  Testing and Analysis (ISSTA)}}. \bibinfo{publisher}{ACM},
  \bibinfo{pages}{682--685}.
\newblock


\bibitem[McMinn(2004)]%
        {Mcm04}
\bibfield{author}{\bibinfo{person}{P. McMinn}.}
  \bibinfo{year}{2004}\natexlab{}.
\newblock \showarticletitle{Search-based Software Test Data Generation: A
  Survey}.
\newblock \bibinfo{journal}{\emph{Software Testing, Verification and
  Reliability}} \bibinfo{volume}{14}, \bibinfo{number}{2}
  (\bibinfo{year}{2004}), \bibinfo{pages}{105--156}.
\newblock


\bibitem[Newman(2021)]%
        {newman2021building}
\bibfield{author}{\bibinfo{person}{Sam Newman}.}
  \bibinfo{year}{2021}\natexlab{}.
\newblock \bibinfo{booktitle}{\emph{Building microservices}}.
\newblock \bibinfo{publisher}{" O'Reilly Media, Inc."}.
\newblock


\bibitem[Panichella et~al\mbox{.}(2018)]%
        {dynamosa2017}
\bibfield{author}{\bibinfo{person}{Annibale Panichella},
  \bibinfo{person}{Fitsum Kifetew}, {and} \bibinfo{person}{Paolo Tonella}.}
  \bibinfo{year}{2018}\natexlab{}.
\newblock \showarticletitle{Automated Test Case Generation as a Many-Objective
  Optimisation Problem with Dynamic Selection of the Targets}.
\newblock \bibinfo{journal}{\emph{IEEE Transactions on Software Engineering
  (TSE)}} \bibinfo{volume}{44}, \bibinfo{number}{2} (\bibinfo{year}{2018}),
  \bibinfo{pages}{122--158}.
\newblock


\bibitem[Pitzer and Affenzeller(2012)]%
        {pitzer2012comprehensive}
\bibfield{author}{\bibinfo{person}{Erik Pitzer} {and} \bibinfo{person}{Michael
  Affenzeller}.} \bibinfo{year}{2012}\natexlab{}.
\newblock \showarticletitle{A comprehensive survey on fitness landscape
  analysis}.
\newblock \bibinfo{journal}{\emph{Recent advances in intelligent engineering
  systems}} (\bibinfo{year}{2012}), \bibinfo{pages}{161--191}.
\newblock


\bibitem[Poth et~al\mbox{.}(2025)]%
        {poth2025technology}
\bibfield{author}{\bibinfo{person}{Alexander Poth}, \bibinfo{person}{Olsi
  Rrjolli}, {and} \bibinfo{person}{Andrea Arcuri}.}
  \bibinfo{year}{2025}\natexlab{}.
\newblock \showarticletitle{Technology adoption performance evaluation applied
  to testing industrial REST APIs}.
\newblock \bibinfo{journal}{\emph{Automated Software Engineering}}
  \bibinfo{volume}{32}, \bibinfo{number}{1} (\bibinfo{year}{2025}),
  \bibinfo{pages}{5}.
\newblock


\bibitem[Qui{\~n}a-Mera et~al\mbox{.}(2023)]%
        {quina2023graphql}
\bibfield{author}{\bibinfo{person}{Antonio Qui{\~n}a-Mera},
  \bibinfo{person}{Pablo Fernandez}, \bibinfo{person}{Jos{\'e}~Mar{\'\i}a
  Garc{\'\i}a}, {and} \bibinfo{person}{Antonio Ruiz-Cort{\'e}s}.}
  \bibinfo{year}{2023}\natexlab{}.
\newblock \showarticletitle{GraphQL: A Systematic Mapping Study}.
\newblock \bibinfo{journal}{\emph{Comput. Surveys}} \bibinfo{volume}{55},
  \bibinfo{number}{10} (\bibinfo{year}{2023}), \bibinfo{pages}{1--35}.
\newblock


\bibitem[Rajesh(2016)]%
        {rajesh2016spring}
\bibfield{author}{\bibinfo{person}{RV Rajesh}.}
  \bibinfo{year}{2016}\natexlab{}.
\newblock \bibinfo{booktitle}{\emph{Spring Microservices}}.
\newblock \bibinfo{publisher}{Packt Publishing Ltd}.
\newblock


\bibitem[Sahin et~al\mbox{.}(2025)]%
        {sahinFitnessReplication2025}
\bibfield{author}{\bibinfo{person}{Omur Sahin}, \bibinfo{person}{Man Zhang},
  {and} \bibinfo{person}{Andrea Arcuri}.} \bibinfo{year}{2025}\natexlab{}.
\newblock \bibinfo{booktitle}{\emph{Replication Package for Causes and Effects
  of Fitness Landscapes in System Test Generation: A Replication Study}}.
\newblock
\urldef\tempurl%
\url{https://doi.org/10.5281/zenodo.14764981}
\showDOI{\tempurl}


\bibitem[Sayyad et~al\mbox{.}(2013)]%
        {sayyad2013parameter}
\bibfield{author}{\bibinfo{person}{Abdel~Salam Sayyad},
  \bibinfo{person}{Katerina Goseva-Popstojanova}, \bibinfo{person}{Tim
  Menzies}, {and} \bibinfo{person}{Hany Ammar}.}
  \bibinfo{year}{2013}\natexlab{}.
\newblock \showarticletitle{On parameter tuning in search based software
  engineering: A replicated empirical study}. In \bibinfo{booktitle}{\emph{2013
  3rd International Workshop on Replication in Empirical Software Engineering
  Research}}. IEEE, \bibinfo{pages}{84--90}.
\newblock


\bibitem[Shamshiri et~al\mbox{.}(2018)]%
        {shamshiri2018random}
\bibfield{author}{\bibinfo{person}{Sina Shamshiri},
  \bibinfo{person}{Jos{\'e}~Miguel Rojas}, \bibinfo{person}{Luca Gazzola},
  \bibinfo{person}{Gordon Fraser}, \bibinfo{person}{Phil McMinn},
  \bibinfo{person}{Leonardo Mariani}, {and} \bibinfo{person}{Andrea Arcuri}.}
  \bibinfo{year}{2018}\natexlab{}.
\newblock \showarticletitle{Random or evolutionary search for object-oriented
  test suite generation?}
\newblock \bibinfo{journal}{\emph{Software Testing, Verification and
  Reliability}} \bibinfo{volume}{28}, \bibinfo{number}{4}
  (\bibinfo{year}{2018}), \bibinfo{pages}{e1660}.
\newblock


\bibitem[Tawosi et~al\mbox{.}(2021)]%
        {tawosi2021multi}
\bibfield{author}{\bibinfo{person}{Vali Tawosi}, \bibinfo{person}{Federica
  Sarro}, \bibinfo{person}{Alessio Petrozziello}, {and} \bibinfo{person}{Mark
  Harman}.} \bibinfo{year}{2021}\natexlab{}.
\newblock \showarticletitle{Multi-objective software effort estimation: A
  replication study}.
\newblock \bibinfo{journal}{\emph{IEEE Transactions on Software Engineering}}
  \bibinfo{volume}{48}, \bibinfo{number}{8} (\bibinfo{year}{2021}),
  \bibinfo{pages}{3185--3205}.
\newblock


\bibitem[Vassilev et~al\mbox{.}(2000)]%
        {vassilev2000information}
\bibfield{author}{\bibinfo{person}{Vesselin~K Vassilev},
  \bibinfo{person}{Terence~C Fogarty}, {and} \bibinfo{person}{Julian~F
  Miller}.} \bibinfo{year}{2000}\natexlab{}.
\newblock \showarticletitle{Information characteristics and the structure of
  landscapes}.
\newblock \bibinfo{journal}{\emph{Evolutionary computation}}
  \bibinfo{volume}{8}, \bibinfo{number}{1} (\bibinfo{year}{2000}),
  \bibinfo{pages}{31--60}.
\newblock


\bibitem[Viglianisi et~al\mbox{.}(2020)]%
        {viglianisi2020resttestgen}
\bibfield{author}{\bibinfo{person}{Emanuele Viglianisi},
  \bibinfo{person}{Michael Dallago}, {and} \bibinfo{person}{Mariano Ceccato}.}
  \bibinfo{year}{2020}\natexlab{}.
\newblock \showarticletitle{RESTTESTGEN: Automated Black-Box Testing of RESTful
  APIs}. In \bibinfo{booktitle}{\emph{IEEE International Conference on Software
  Testing, Verification and Validation (ICST)}}. IEEE.
\newblock


\bibitem[Vogel et~al\mbox{.}(2019)]%
        {vogel2019does}
\bibfield{author}{\bibinfo{person}{Thomas Vogel}, \bibinfo{person}{Chinh Tran},
  {and} \bibinfo{person}{Lars Grunske}.} \bibinfo{year}{2019}\natexlab{}.
\newblock \showarticletitle{Does diversity improve the test suite generation
  for mobile applications?}. In \bibinfo{booktitle}{\emph{International
  Symposium on Search Based Software Engineering}}. Springer,
  \bibinfo{pages}{58--74}.
\newblock


\bibitem[Vogel et~al\mbox{.}(2021)]%
        {vogel2021comprehensive}
\bibfield{author}{\bibinfo{person}{Thomas Vogel}, \bibinfo{person}{Chinh Tran},
  {and} \bibinfo{person}{Lars Grunske}.} \bibinfo{year}{2021}\natexlab{}.
\newblock \showarticletitle{A comprehensive empirical evaluation of generating
  test suites for mobile applications with diversity}.
\newblock \bibinfo{journal}{\emph{Information and Software Technology}}
  \bibinfo{volume}{130} (\bibinfo{year}{2021}), \bibinfo{pages}{106436}.
\newblock


\bibitem[Waeselynck et~al\mbox{.}(2006)]%
        {WFK06}
\bibfield{author}{\bibinfo{person}{H. Waeselynck}, \bibinfo{person}{P.~T.
  Fosse}, {and} \bibinfo{person}{O.~A. Kaddour}.}
  \bibinfo{year}{2006}\natexlab{}.
\newblock \showarticletitle{Simulated annealing applied to test generation:
  landscape characterization and stopping criteria}.
\newblock \bibinfo{journal}{\emph{Empirical Software Engineering}}
  \bibinfo{volume}{12}, \bibinfo{number}{1} (\bibinfo{year}{2006}),
  \bibinfo{pages}{35--63}.
\newblock


\bibitem[Wu et~al\mbox{.}(2022)]%
        {wu2022icse}
\bibfield{author}{\bibinfo{person}{Huayao Wu}, \bibinfo{person}{Lixin Xu},
  \bibinfo{person}{Xintao Niu}, {and} \bibinfo{person}{Changhai Nie}.}
  \bibinfo{year}{2022}\natexlab{}.
\newblock \showarticletitle{Combinatorial Testing of RESTful APIs}. In
  \bibinfo{booktitle}{\emph{ACM/IEEE International Conference on Software
  Engineering (ICSE)}}.
\newblock


\bibitem[Zhang and Arcuri(2021a)]%
        {zhang2021adaptive}
\bibfield{author}{\bibinfo{person}{Man Zhang} {and} \bibinfo{person}{Andrea
  Arcuri}.} \bibinfo{year}{2021}\natexlab{a}.
\newblock \showarticletitle{Adaptive Hypermutation for Search-Based System Test
  Generation: A Study on REST APIs with EvoMaster}.
\newblock \bibinfo{journal}{\emph{ACM Transactions on Software Engineering and
  Methodology (TOSEM)}} \bibinfo{volume}{31}, \bibinfo{number}{1}
  (\bibinfo{year}{2021}).
\newblock


\bibitem[Zhang and Arcuri(2021b)]%
        {zhang2021enhancing}
\bibfield{author}{\bibinfo{person}{Man Zhang} {and} \bibinfo{person}{Andrea
  Arcuri}.} \bibinfo{year}{2021}\natexlab{b}.
\newblock \showarticletitle{Enhancing Resource-Based Test Case Generation for
  RESTful APIs with SQL Handling}. In \bibinfo{booktitle}{\emph{International
  Symposium on Search Based Software Engineering}}. Springer,
  \bibinfo{pages}{103--117}.
\newblock


\bibitem[Zhang and Arcuri(2023)]%
        {zhang2023open}
\bibfield{author}{\bibinfo{person}{Man Zhang} {and} \bibinfo{person}{Andrea
  Arcuri}.} \bibinfo{year}{2023}\natexlab{}.
\newblock \showarticletitle{Open Problems in Fuzzing RESTful APIs: A Comparison
  of Tools}.
\newblock  (\bibinfo{year}{2023}).
\newblock
\showISSN{1049-331X}
\urldef\tempurl%
\url{https://doi.org/10.1145/3597205}
\showDOI{\tempurl}


\bibitem[Zhang et~al\mbox{.}(2023a)]%
        {zhang2023rpc}
\bibfield{author}{\bibinfo{person}{Man Zhang}, \bibinfo{person}{Andrea Arcuri},
  \bibinfo{person}{Yonggang Li}, \bibinfo{person}{Yang Liu}, {and}
  \bibinfo{person}{Kaiming Xue}.} \bibinfo{year}{2023}\natexlab{a}.
\newblock \showarticletitle{White-box fuzzing rpc-based apis with evomaster: An
  industrial case study}.
\newblock \bibinfo{journal}{\emph{ACM Transactions on Software Engineering and
  Methodology}} \bibinfo{volume}{32}, \bibinfo{number}{5}
  (\bibinfo{year}{2023}), \bibinfo{pages}{1--38}.
\newblock


\bibitem[Zhang et~al\mbox{.}(2024)]%
        {zhang2024seeding}
\bibfield{author}{\bibinfo{person}{Man Zhang}, \bibinfo{person}{Andrea Arcuri},
  \bibinfo{person}{Piyun Teng}, \bibinfo{person}{Kaiming Xue}, {and}
  \bibinfo{person}{Wenhao Wang}.} \bibinfo{year}{2024}\natexlab{}.
\newblock \showarticletitle{Seeding and Mocking in White-Box Fuzzing Enterprise
  RPC APIs: An Industrial Case Study}. In \bibinfo{booktitle}{\emph{Proceedings
  of the 39th IEEE/ACM International Conference on Automated Software
  Engineering}}. \bibinfo{pages}{2024--2034}.
\newblock


\bibitem[Zhang et~al\mbox{.}(2023b)]%
        {zhang2023javascript}
\bibfield{author}{\bibinfo{person}{Man Zhang}, \bibinfo{person}{Asma Belhadi},
  {and} \bibinfo{person}{Andrea Arcuri}.} \bibinfo{year}{2023}\natexlab{b}.
\newblock \showarticletitle{JavaScript SBST Heuristics To Enable Effective
  Fuzzing of NodeJS Web APIs}.
\newblock \bibinfo{journal}{\emph{ACM Transactions on Software Engineering and
  Methodology}} (\bibinfo{year}{2023}).
\newblock


\bibitem[Zhang et~al\mbox{.}(2021)]%
        {zhang2021resource}
\bibfield{author}{\bibinfo{person}{Man Zhang}, \bibinfo{person}{Bogdan
  Marculescu}, {and} \bibinfo{person}{Andrea Arcuri}.}
  \bibinfo{year}{2021}\natexlab{}.
\newblock \showarticletitle{Resource and dependency based test case generation
  for RESTful Web services}.
\newblock \bibinfo{journal}{\emph{Empirical Software Engineering}}
  \bibinfo{volume}{26}, \bibinfo{number}{4} (\bibinfo{year}{2021}),
  \bibinfo{pages}{1--61}.
\newblock


\bibitem[Zou et~al\mbox{.}(2022)]%
        {zou2022survey}
\bibfield{author}{\bibinfo{person}{Feng Zou}, \bibinfo{person}{Debao Chen},
  \bibinfo{person}{Hui Liu}, \bibinfo{person}{Siyu Cao},
  \bibinfo{person}{Xuying Ji}, {and} \bibinfo{person}{Yan Zhang}.}
  \bibinfo{year}{2022}\natexlab{}.
\newblock \showarticletitle{A survey of fitness landscape analysis for
  optimization}.
\newblock \bibinfo{journal}{\emph{Neurocomputing}}  \bibinfo{volume}{503}
  (\bibinfo{date}{9} \bibinfo{year}{2022}), \bibinfo{pages}{129--139}.
\newblock
\showISSN{0925-2312}
\urldef\tempurl%
\url{https://doi.org/10.1016/J.NEUCOM.2022.06.084}
\showDOI{\tempurl}


\end{thebibliography}

%If needed
%\input{appendix.tex}

\end{document}